\documentclass[12pt, preprint]{aastex}

\newcommand{\map}{\textsl{WMAP}}

\newcommand{\WMAP}{\textsl{WMAP}}

\newcommand{\Brown}{{Department of Physics, Brown University, %
            182 Hope St., Providence, RI 02912-1843, USA}}

\providecommand{\degrees}{\ensuremath{{}^{\circ}}}
\providecommand{\gt}{\ensuremath{>}}
\providecommand{\muK}{\ensuremath{\;\mu\textrm{K}}}

\shorttitle{\WMAP\ Seven-Year Galactic Foreground Emission}
\shortauthors{Gold et al.}

\begin{document}
\title{Seven-Year {\sl Wilkinson Microwave Anisotropy Probe (\WMAP\altaffilmark{1})} Observations: \\
Galactic Foreground Emission}
\author{B.~Gold\altaffilmark{2}, 
N.~Odegard\altaffilmark{3},
J.~L.~Weiland\altaffilmark{3},
R.~S.~Hill\altaffilmark{3},
A.~Kogut\altaffilmark{4}, 
C.~L.~Bennett\altaffilmark{2},
G.~Hinshaw\altaffilmark{4}, 
X.~Chen\altaffilmark{5},
J.~Dunkley\altaffilmark{6},
M.~Halpern\altaffilmark{7},
N.~Jarosik\altaffilmark{8}, 
E.~Komatsu\altaffilmark{9}, 
D.~Larson\altaffilmark{2}, 
M.~Limon\altaffilmark{10},
S.~S.~Meyer\altaffilmark{11},
M.~R.~Nolta\altaffilmark{12}, 
L.~Page\altaffilmark{8},
K.~M.~Smith\altaffilmark{13},
D.~N.~Spergel\altaffilmark{8,13},
G.~S.~Tucker\altaffilmark{14},
E.~Wollack\altaffilmark{4}, 
and E.~L.~Wright\altaffilmark{15}}
%\author{
%B. Gold\altaffilmark{2},
%C. L. Bennett\altaffilmark{2},
%R. S. Hill\altaffilmark{3}, 
%G. Hinshaw\altaffilmark{4}, 
%N. Odegard\altaffilmark{3}, 
%L. Page\altaffilmark{5}, 
%D. N. Spergel\altaffilmark{6,7}, 
%J. L. Weiland\altaffilmark{3},   
%J. Dunkley\altaffilmark{5,7,8}, 
%M. Halpern\altaffilmark{9}, 
%N. Jarosik\altaffilmark{5}, 
%A. Kogut\altaffilmark{4}, 
%E. Komatsu\altaffilmark{10}, 
%D. Larson\altaffilmark{2}, 
%S. S. Meyer\altaffilmark{11},
%M. R. Nolta\altaffilmark{12}, 
%E. Wollack\altaffilmark{4}, 
%and E. L. Wright\altaffilmark{13}
%}
\altaffiltext{1}{{\WMAP} is the result of a partnership between 
                 Princeton  University and NASA's Goddard Space Flight Center. 
                 Scientific guidance is provided by the 
                 {\WMAP} Science Team.}
\altaffiltext{2}{Department of Physics \& Astronomy, %
            The Johns Hopkins University, 3400 N. Charles St., %
	    Baltimore, MD  21218-2686, USA}
\altaffiltext{3}{Adnet Systems, Inc., 7515 Mission Dr., Suite A1C1 Lanham, MD 20706, USA}
\altaffiltext{4}{Code 665, NASA/Goddard Space Flight Center, %
            Greenbelt, MD 20771, USA}
\altaffiltext{5}{Infrared Processing and Analysis Center, California Institute of Technology, 1200 E. California Blvd., Pasadena, CA 91125, USA}
\altaffiltext{6}{Astrophysics, University of Oxford, Keble Road, Oxford, OX1 3RH, UK}
\altaffiltext{7}{Department of Physics and Astronomy, University of %
            British Columbia, Vancouver, BC  Canada V6T 1Z1}
\altaffiltext{8}{Department of Physics, Jadwin Hall, %
            Princeton University, Princeton, NJ 08544-0708, USA}
\altaffiltext{9}{University of Texas, Austin, Department of Astronomy, %
            2511 Speedway, RLM 15.306, Austin, TX 78712, USA}
\altaffiltext{10}{Columbia Astrophysics Lab, Columbia University, %
            Mail Code 5247, 550 W. 120th St, New York, NY 10027, USA}
\altaffiltext{11}{Departments of Astrophysics and Physics, KICP and EFI, %
            University of Chicago, Chicago, IL 60637, USA}
\altaffiltext{12}{Canadian Institute for Theoretical Astrophysics, %
            60 St. George St, University of Toronto, %
	    Toronto, ON  Canada M5S 3H8}
\altaffiltext{13}{Department of Astrophysical Sciences, %
            Peyton Hall, Princeton University, Princeton, NJ 08544-1001, USA}
%\altaffiltext{7}{Princeton Center for Theoretical Physics, Princeton University, Princeton, NJ 08544}
\altaffiltext{14}{\Brown}
\altaffiltext{15}{PAB 3-909, UCLA Physics \& Astronomy, PO Box 951547, %
            Los Angeles, CA 90095--1547, USA}
\setcounter{footnote}{15}

\email{bgold@pha.jhu.edu}

\begin{abstract}
We present updated estimates of Galactic foreground emission using seven years of \WMAP\ data.  Using the power spectrum of differences between multi-frequency template-cleaned maps, we find no evidence for foreground contamination outside of the updated (KQ85y7) foreground mask.  We place a 15 $\mu$K upper bound on {rms} foreground contamination in the cleaned maps used for cosmological analysis.
Further, the cleaning process requires only three power-law foregrounds outside of the mask.  We find no evidence for polarized foregrounds beyond those from soft (steep-spectrum) synchrotron and thermal dust emission; in particular we find no indication in the polarization data of an extra ``haze'' of  hard synchrotron emission from energetic electrons near the Galactic center.
We provide an updated map of the cosmic microwave background (CMB) using the internal linear combination method, updated foreground masks, and updates to point source catalogs using two different techniques.  With additional years of data, we now detect 471 point sources using a five-band technique and 417 sources using a three-band CMB-free technique.  In total there are 62 newly detected point sources, a 12\% increase over the five-year release.  Also new are tests of the Markov chain Monte Carlo foreground fitting procedure against systematics in the time-stream data, and tests against the observed beam asymmetry.

Within a few degrees of the Galactic plane, the behavior in total intensity of low-frequency foregrounds is complicated and not completely understood.  \WMAP\ data show a rapidly steepening spectrum from 20-40 GHz, which may be due to emission from spinning dust grains, steepening synchrotron, or other effects.  Comparisons are made to a 1-degree 408 MHz map (Haslam et al.) and the 11-degree ARCADE 2 data (Singal et al.).  We find that spinning dust or steepening synchrotron models fit the combination of \WMAP\ and 408 MHz data equally well.  ARCADE data appear inconsistent with the steepening synchrotron model, and consistent with the spinning dust model, though some discrepancies remain regarding the relative strength of spinning dust emission.  More high-resolution data in the 10-40 GHz range would shed much light on these issues.
\end{abstract}

%\slugcomment{Draft date: \today}
%\slugcomment{submitted to ApJS}
%\slugcomment{Revised version, accepted for publication by ApJS}
\slugcomment{Published in ApJS vol 191}
\maketitle

% for organizational purposes, not for final display
%\tableofcontents

%%% BEGINNING OF TEXT %%%%%%%%%%%%%%%%%%%%%%%%%%%%%%
%%% Introduction %%%%%%%%%%%%%%%%%%%%%%%%%%%%
%%%%%%%%%%%%%%%%%%%%%%%%%%%%%%%%%%%%%%%%%%%
\section{Introduction}

The {\it Wilkinson Microwave Anisotropy Probe} (\WMAP) was launched in 2001 to observe the cosmic microwave background (CMB).
In addition to measuring the CMB, \WMAP, like any full-sky CMB experiment, also observes emission from our own Galaxy. 
With five frequency bands centered at 23, 33, 41, 61, and 94 GHz (respectively denoted as $K$, $Ka$, $Q,$ $V$, and $W$ bands), full sky coverage, polarization sensitivity, and control of systematics to the sub-percent level,  
\WMAP\ is able to measure diffuse ($1\degrees$ and larger) emission with precise temperature calibration.  In this paper we analyze seven years of \WMAP\ data in order to better characterize Galactic foreground emission, the removal of which will be one of the largest challenges to future CMB experiments \citep{dunkley/etal:2009c}.

This paper is part of a suite of papers describing the full details of the \WMAP\ seven-year data release.  An overall description of sky maps and basic results is in \cite{jarosik/etal:prep}, which also includes a description of the beam modeling used to produce maps smoothed to the common resolution of a $1\degrees$ FWHM Gaussian.  These maps serve as the starting point for foreground analysis in this work.  \cite{larson/etal:prep} describe the generation of power spectra from CMB maps, and \cite{komatsu/etal:prep} discuss the cosmological implications of the spectra.  \cite{weiland/etal:prep} detail measurements of celestial calibrators, and \cite{bennett/etal:prep} investigate the status of some potential anomalies found in \WMAP\ data. 

The layout of this paper is as follows.  Updates to masks and foreground fitting processes are described in Section~\ref{sec:sevenyearfits}.  A comparison of \WMAP\ data to that recently taken by the ARCADE instrument \citep{singal/etal:2010} is discussed in Section~\ref{sec:arcade}.  Results of the fits and their implications for specific foreground emission processes are discussed in Section~\ref{sec:results}.  
A discussion of systematics follows in Section~\ref{sec:tests}.  Point sources and an update to the point source catalog are found in Section~\ref{sec:pointsources}.  Lastly, conclusions can be found in Section~\ref{sec:conclusions}.

\subsection{Science Overview}

There are three primary mechanisms for diffuse Galactic radio emission. Relativistic electrons interact with the Galactic magnetic field to produce synchrotron emission, for which the standard template is 408 MHz data compiled by \cite{haslam/etal:1981}.  Less energetic electrons scatter from each other and ionized nuclei to produce free-free radiation (also known as thermal Bremsstrahlung), which can be traced with H$\alpha$ line emission \citep{finkbeiner:2003}.  Finally, dust grains emit a modified blackbody spectrum through excitation of their vibrational modes, for which the standard template is the fit of \cite{finkbeiner/davis/schlegel:1999} to data from the {\it Infrared Astronomical Satellite (IRAS)} and the {\it Cosmic Background Explorer (COBE)}. Dust grains may also emit radiation through rotational modes or other excitations \citep{draine/lazarian:1998a, draine/lazarian:1998b, draine/lazarian:1999}.

\WMAP\ was designed to measure near the frequency where the ratio of the CMB anisotropy to the rms fluctuations of all three foregrounds is at its maximum, to minimize foreground contamination.  This also implies that two or more foreground components will be of comparable amplitude and that they will be relatively weak. 
Foreground templates, however, are best made by observing a foreground process at a
frequency where it dominates the total emission.  Hence there will always be
 some extrapolation involved when attempting to account for foregrounds on
top of CMB observations.  

So how well does the extrapolation work?  Simple
power-law extrapolation of the 408 MHz synchrotron template from \cite{haslam/etal:1981} does
not explain very much of the observed emission at 20-40 GHz.  Whether this is due to a
new low-frequency emission process, errors in the extrapolation due to spatial variation in the spectral index, or
both, is difficult to determine.  
Targeted observations of individual regions \citep{scaife/etal:2010, tibbs/etal:2010} suggest a spinning dust-like component, but a model consistent across size scales and data sets remains elusive.

Free-free emission is extrapolated from a map of H$\alpha$, corrected for dust
extinction using a reddening map based on 100 $\mu$m data \citep{schlegel/finkbeiner/davis:1998, bennett/etal:2003c}.  Variations in
electron temperature cause some uncertainty in this extrapolation, but the
larger effect is likely uncertainty in the reddening correction.  The overall ratio of radio to H$\alpha$ brightness comes out lower than expected \citep{bennett/etal:2003c}; nevertheless,
the template otherwise matches quite well with observations at 30-60 GHz, where
free-free emission from the Galactic disk is particularly dominant.

The dust extrapolation has so far been tested least precisely by CMB experiments.  While the
model of \cite{finkbeiner/davis/schlegel:1999} incorporates {\it COBE} FIRAS data all the way 
down to 60 GHz,
the uncertainty at those frequencies is large; most of the dust model
comes from information at 100 and 240 $\mu$m.  
While the spectral index of dust at frequencies below 300 GHz has not yet been measured to enough accuracy to challenge the model, 
the morphology matches observations at lower frequencies, though
some experiments suggest overall brightness levels different from the predictions
\citep{veneziani/etal:2009, culverhouse/etal:2010}.

Analysis of data from previous \WMAP\ releases has shown that CMB maps from different foreground removal techniques agree to within 11 $\mu$K \citep{gold/etal:2009} on average in the low Galactic emission regions used for CMB anisotropy measurements, though this does not provide an absolute limit to the amount of contamination.
Even when templates are not directly used for foreground removal, they provide an important guide for the construction of masks and other foreground cleaning methods.

Several systems of units are in use throughout this work.  Point sources are reported in flux units (Jansky), where the power-law index is denoted by $\alpha$ such that flux follows $S \sim \nu^\alpha$.  Foreground modeling is most easily done in units of antenna temperature, defined by using the Rayleigh-Jeans limit of a blackbody spectrum (for which $S \sim \nu^2$) to convert flux per solid angle to a temperature.  In these units the power-law index is denoted as $\beta = \alpha-2$.  \WMAP's frequency range is not quite in the Rayleigh-Jeans limit for a $2.7$ K blackbody, so there is a frequency-dependent conversion factor $a(\nu) = (e^x-1)^2/x^2e^x$ (where $x=h\nu/k_B T_\mathrm{cmb}$) to convert antenna temperatures to thermodynamic temperatures convenient for CMB analysis. 

%%% 7yr UPDATE %%%%%%%%%%%%%%%%%%%%%%%%%%%%
\section{Seven-year Foreground Fits\label{sec:sevenyearfits}}

\subsection{Masks}
Foreground removal always has some uncertainty, so it is useful to mask part of the sky where foregrounds are too bright for CMB analysis. As in the five-year analysis, the starting points for the masks are $K$- and $Q$-band-average maps smoothed to one-degree resolution.  The maps are then converted to foreground-only maps by
subtracting off an estimate of the CMB using the internal linear combination (ILC) method (see \citealt{hinshaw/etal:2007}, and Section~\ref{ssec:ilc}).  A cumulative histogram in each band is formed to find the flux level above which a given percentage of sky can be cut, and the union of the pixels cut from each band at a given flux level is used to define a mask.  We used two masks for most further analysis, based on cuts which leave 75\% and 85\% of the sky; these are denoted as KQ75 and KQ85, respectively.

For the seven-year analysis, the diffuse foreground masks have been extended based on a $\chi^2$ analysis of residuals after foreground subtraction.  Starting with foreground-reduced maps, differences are taken between bands ($Q-V$ and $V-W$ in thermodynamic units), eliminating any CMB signal.  Ideally the only thing left in the resulting maps would be noise; in practice there are visible residuals near the Galactic plane.
%, as the foreground characterization begins to break down. 
Given the noise per pixel of the maps, it is possible to compute a map of the $\chi^2$ for each pixel.

After degradation to HEALPix $N_\mathrm{side} = 32$ (see \citealt{gorski/etal:2005} for a description of this pixelization scheme), regions of 4 or more contiguous pixels with $\chi^2$ higher than four times that of the polar caps are identified and used to define two new masks, one from each difference map.  These are then combined with the previous KQ75 and KQ85 masks \citep{gold/etal:2009} used for the five-year analysis.  After promotion back to full resolution, edges of the mask are smoothed with a $3\degrees$ FWHM Gaussian.  The resulting changes to the final mask are primarily around the edge of the Galactic cut, particularly in the Gum and Ophiuchus regions.  The additional sky fraction cut from the KQ85 masked sky is $3.4\%$, and from the KQ75 masked sky is $1.0\%$.

\begin{figure}
\epsscale{0.7}
%\plotone{compare_masks_kq75y7y5.pdf}
%\plotone{compare_masks_kq85y7y5_key.pdf}
%\plotone{f01a.pdf}
%\plotone{f01b.pdf}
\plotone{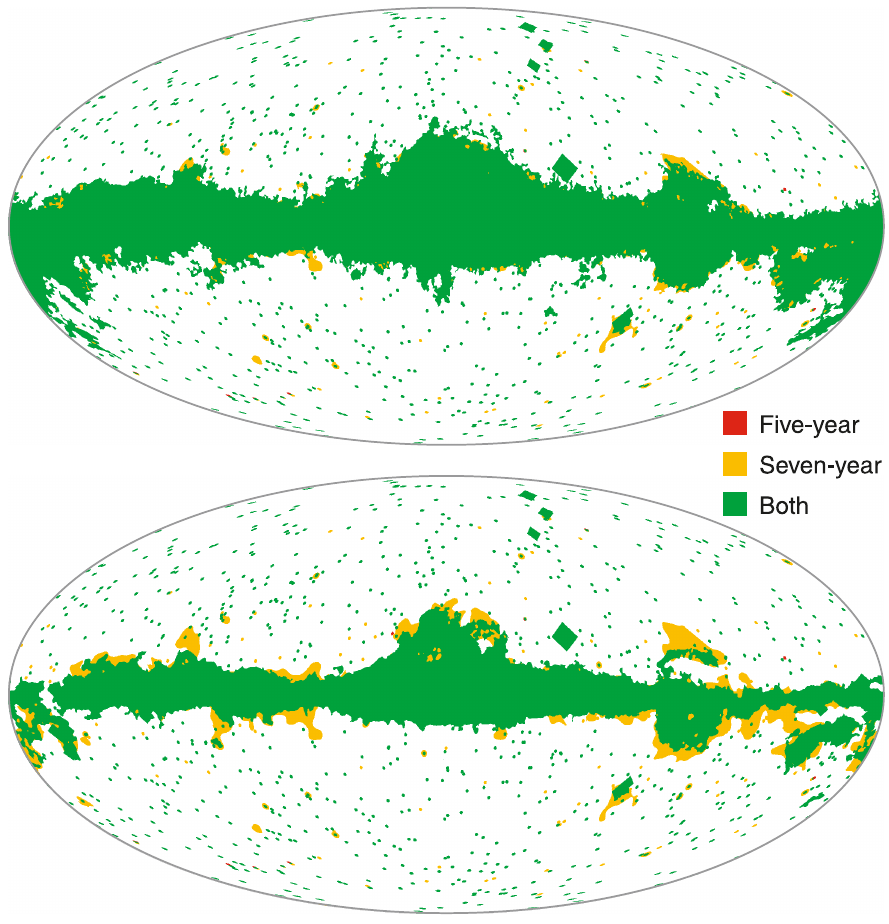}
\caption{\label{fig:maskmap}
Comparison of seven-year masks to five-year masks.  At the top KQ75 and KQ75y7 are compared, and at the bottom KQ85 and KQ85y7.  Green regions are masked in both the seven-year and five-year masks, yellow regions are newly masked in the seven-year masks, and red regions are masked in the five-year masks but no longer in the seven-year masks.
}
\end{figure}

These expanded masks are then combined with the point source mask as in previous releases, which has been updated with newly detected sources.  Also, point sources brighter than $5$ Jy have had the radius of their cut extended from $0\degrees.6$ to $1\degrees.2$, in order to minimize confusion at low frequencies where the instrument beam is large.  The new masks, which we denote KQ75y7 and KQ85y7, are shown in Figure \ref{fig:maskmap}, and are available on the LAMBDA Web site\footnote{\texttt{http://lambda.gsfc.nasa.gov/}}.  In total $70.6\%$ (KQ75y7) and  $78.3\%$ (KQ85y7) of the sky now remains after the masking process.

\subsection{Internal Linear Combination Method\label{ssec:ilc}}
%The Internal Linear Combination (ILC) method implemented by \WMAP\ is a remarkably robust technique, blind to assumptions about the spectrum of foreground emission, 
%which produces CMB maps with little visible foreground contamination. The variance between the ILC map and CMB maps made with other techniques is less than 116 $\mu\textrm{K}^2$ \citep{gold/etal:2009}.
The ILC method implemented by \WMAP\ is a technique largely blind to assumptions about the frequency spectrum of foreground emission, which produces CMB maps with little visible foreground contamination.  
The ILC is a weighted combination formed from all five frequency bands, which are smoothed to a common $1\degrees$ FWHM Gaussian beam using the symmetrized beam window functions produced by the beam analysis \citep{jarosik/etal:prep}.  The coefficients used to weight each individual frequency band are those that minimize the variance of the resulting map under the constraint that the sum of the coefficients is unity, which ensures that the CMB portion of the signal is passed through unaltered.

The details of the algorithm used to compute the \WMAP\ seven-year ILC map are the same as that described in the three-year analysis
\citep{hinshaw/etal:2007}.
In particular, we perform a bias correction step which uses simulations to estimate and correct for the tendency of the ILC method to produce CMB maps anti-correlated with foreground fluctuations (for an overview of potential ILC pitfalls see \citealt{vio/andreani:2009}).
We have found this technique to be robust when applied to \WMAP\ data: the variance between the ILC map and CMB maps made with other techniques is less than 116 $\mu\textrm{K}^2$ \citep{gold/etal:2009}.  Similar techniques by other authors have given CMB maps consistent with \WMAP's best-fit cosmological results \citep{kim/naselsky/christensen:2008}.

\begin{figure}
\epsscale{1.0}
%\plotone{ilc75diff.pdf}
\plotone{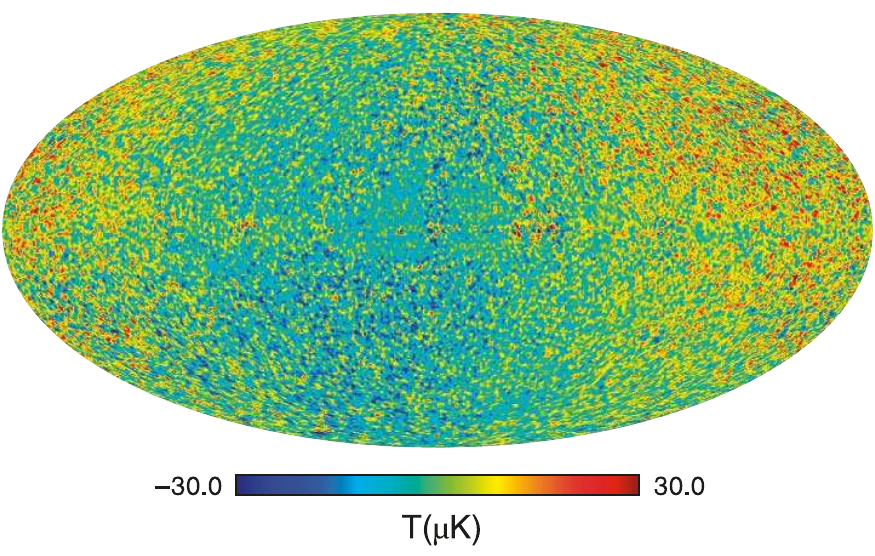}
\caption{\label{fig:ilcdiff}
Difference map between the seven-year ILC map and the five-year ILC map.  Small-scale differences are consistent with pixel noise; large-scale differences are consistent with a change in dipole of $6.7$ $\mu\textrm{K}$.
}
\end{figure}

Rather than use a single set of coefficients for the whole sky, to allow for variations in Galactic composition we subdivide the sky into 12 regions and find the ILC coefficients for each, shown in Table~\ref{tab:ilc}.  All but region 0 lie along the Galactic plane.
We retain the same number of regional subdivisions
of the sky and their spatial boundaries remain unchanged from the
previous years (for details see \citealt{hinshaw/etal:2007}). The frequency weights for each region are slightly
different, however, reflecting the most recent updates to the calibration and beams.  Figure~\ref{fig:ilcdiff} shows the difference between the seven-year and five-year ILC maps, which is dominated by a small change in the dipole.  The seven-year ILC map is available on the LAMBDA Web site.

\subsection{Maximum Entropy Method}
The maximum entropy method (MEM) is a spatial and spectral fit that uses external templates, intended to distinguish between different emission sources.  By design, the MEM output tends to revert to these templates in regions of low signal-to-noise.  Thus the MEM results are most interesting in regions with higher signal.

The seven-year MEM analysis is largely unchanged from previous work \citep{hinshaw/etal:2007, gold/etal:2009}.  As before, the analysis is done in all bands on the $1\degrees$ smoothed sky maps, with the ILC map subtracted.  The zero level of each map is set such that a $\csc |b|$ fit, for HEALPix $N_\mathrm{side} = 512$ pixels at $b < -15\degrees$ and outside of the KQ85y7 mask, yields a value of zero for the intercept.
The maps are degraded to HEALPix $N_\mathrm{side} = 128$ pixelization, and a model is fit for each pixel $p$.  Rather than simply minimize $\chi^2$, the MEM minimizes a function 
\begin{equation}
   H = \chi^2 + \lambda \sum_c T_c(p) \ln\left[\frac{T_c(p)}{eP_c(p)}\right].
\end{equation}
Here $T_c$ and $P_c$ are the model brightness and template prior for foreground component $c$ ($e$ is the base of natural logarithms). The second term is what enforces the prior when the signal-to-noise becomes low, and the parameter $\lambda$ sets the threshold for the transition from signal-dominated to noise-dominated behavior.  The spectra of the
free-free and dust components are fixed power laws, with $\beta = -2.14$ for free-free and $\beta = +2.0$ for dust.  An iterative procedure uses residuals from the fit at
each iteration to adjust the spectrum of the synchrotron component for each pixel. Hence any anomalous component such as electric dipole emission from spinning dust is included
in the synchrotron component.  The adopted priors are unchanged from previous analyses and are based on the 408 MHz map of \cite{haslam/etal:1981} with an extragalactic brightness of 5.9 K subtracted \citep{lawson/etal:1987} for synchrotron, an
extinction-corrected H$\alpha$ map \citep{finkbeiner:2003, bennett/etal:2003c} for free-free, and model 8 of \cite{finkbeiner/davis/schlegel:1999} for dust.

\begin{figure}
\epsscale{0.8}
\plotone{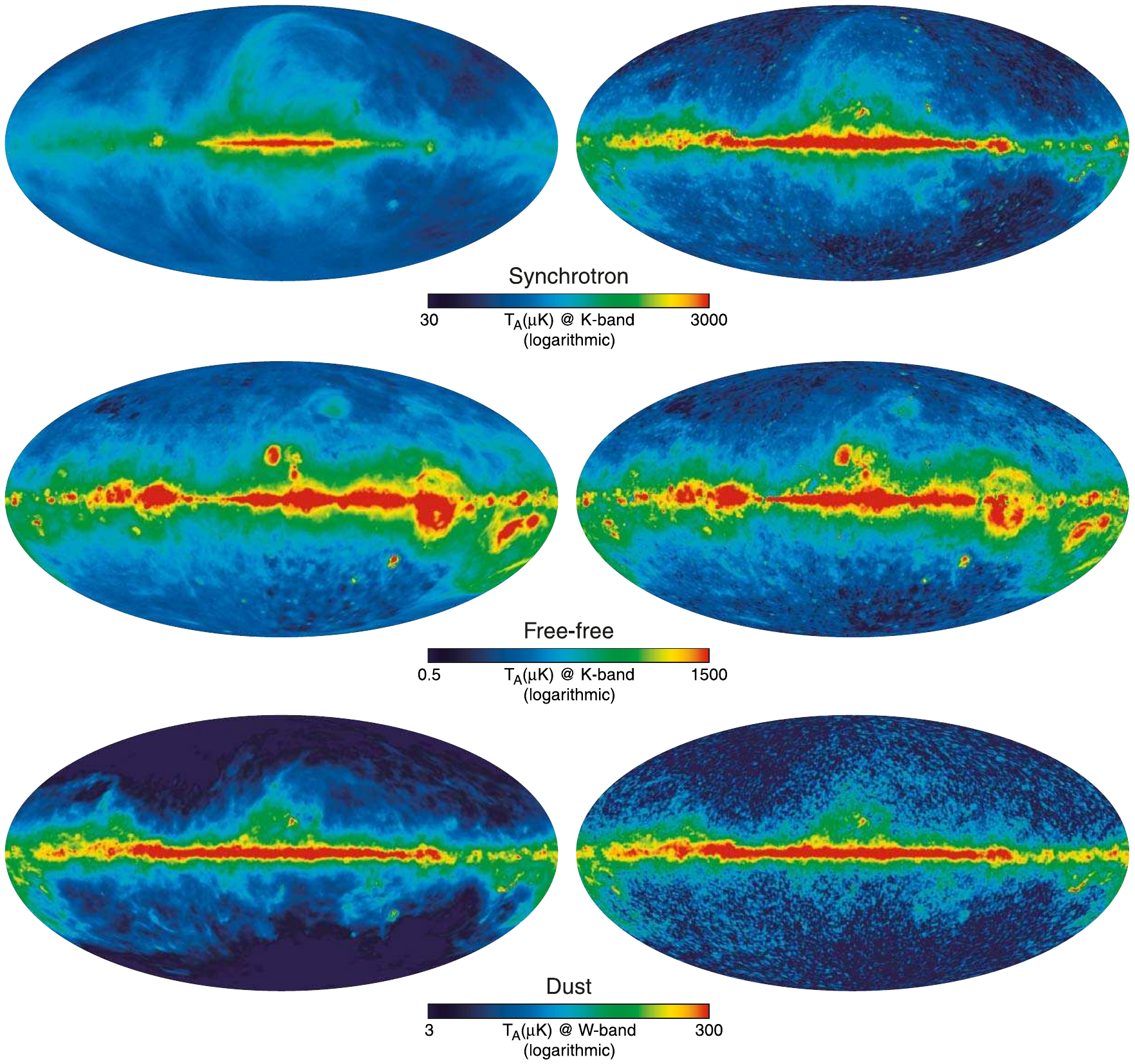}
\caption{\label{fig:memprior}
Galactic signal component maps as determined by the Maximum Entropy Method (MEM) analysis.  On the left are the input prior maps, and on the right are the output MEM maps.  From top to bottom are the synchrotron, free-free, and dust components.  While the output maps show many features of the prior at higher latitudes, there are clear differences in regions of strong emission.
}
\end{figure}

The prior map and output map are shown in Figure~\ref{fig:memprior} for each foreground component.  The zero level of the output synchrotron map is slightly lower
($\sim50 \muK$) than that of the synchrotron prior.  This reflects the difference between the zero level of the $K$-band $\csc|b|$ normalized map and that of the prior.  For
comparison, the $1\sigma$ uncertainty in the prior zero level, based on the quoted uncertainty in the 408 MHz map zero level, is $27 \muK$.  Also, there is a dependence on
the adopted extragalactic brightness at 408 MHz.  If the ARCADE 2 value (\citealt{fixsen/etal:2010}) were used, the zero level of the prior would be $\sim37 \muK$ below that of
the $\csc|b|$ normalized map.  Figure~\ref{fig:memprior} can be compared with Figure~5 of
\citep{hinshaw/etal:2007} to see the improvement in signal-to-noise ratio of the output maps between the three-year and seven-year analyses.

\begin{figure}
\epsscale{0.6}
%\plotone{memdiff2.pdf}
\plotone{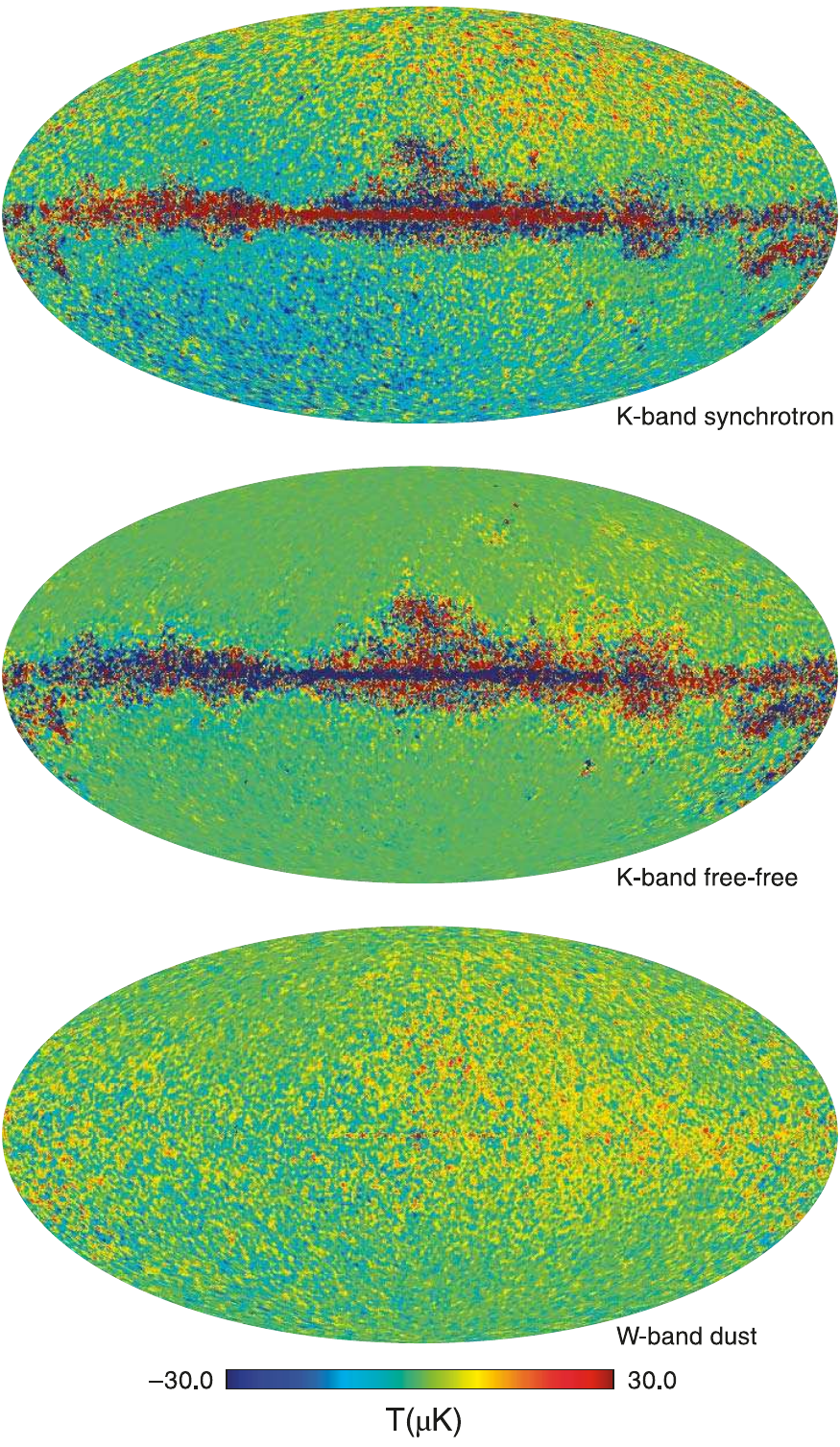}
\caption{\label{fig:memdiff}
Difference maps between the seven-year MEM foreground maps and the five-year MEM foreground maps.  Apart from a small dipole shift and noise fluctuations, the only visible feature is a  small shift of $0.17\%$ of $K$-band flux from free-free to synchrotron.
}
\end{figure}

Differences between seven-year MEM maps and five-year MEM maps are shown in Figure~\ref{fig:memdiff}.
The seven-year MEM foreground component maps tend to be slightly brighter
than the five-year versions at mid to high northern Galactic latitudes.
This is due to small dipole differences between the seven-year and five-year
sky maps, which are caused by a combination of a change in the calibration
dipole and small (less than 0.2\%) changes in radiometer calibrations between the
seven-year and five-year analyses.  The seven-year and five-year foreground    component maps are in better agreement at southern Galactic latitudes because
this is where zero level normalization of the sky maps is determined by $\csc |b|$
fitting. The MEM maps are available on the LAMBDA Web site.

\subsection{Template Cleaning}
\WMAP\ continues to use a template cleaning method to produce the foreground-reduced maps used for power spectrum analysis \citep{hinshaw/etal:2007, page/etal:2007}.  For temperature maps, the templates are a K$-$Ka difference map, an extinction-corrected H$\alpha$ map, and a dust map \cite{finkbeiner/davis/schlegel:1999}.  For polarization, the templates are the $K$-band map for synchrotron, and a dust model described in detail below.

The temperature cleaning is applied to seven-year $Q$-, $V$-, and $W$-band maps ($K$ and $Ka$ are used for a template).  The model has the form
\begin{equation}
   M(\nu,p) = b_1(\nu)\left[T_{K}(p) - T_{Ka}(p)\right] + b_2(\nu) I_\mathrm{H\alpha}(p) 
   + b_3(\nu) M_\mathrm{dust}(p)
\end{equation}
where $p$ indicates the pixel, the frequency dependence is entirely contained in the coefficients $b_i$, 
and the spatial templates are the \WMAP\ K-Ka temperature difference map ($T_{K} - T_{Ka}$), the \cite{finkbeiner:2003}
composite H$\alpha$ map with an extinction correction applied ($I_\mathrm{H\alpha}$), and the
\cite{finkbeiner/davis/schlegel:1999} dust model evaluated at 94 GHz ($M_\mathrm{dust}$).  
Because the first template has contributions from both synchrotron and free-free emission, foreground parameters are a mixture of $b_1(\nu)$ and $b_2(\nu)$.  For free-free emission, the ratio of $K$-band radio temperature to H$\alpha$ intensity is
\begin{equation}
h_\mathrm{ff} = \frac{b_2(\nu)}{S_\mathrm{ff}(\nu) - 0.552\, b_1(\nu)}
\end{equation}
where $S_\mathrm{ff}(\nu)$ is the free-free emission spectrum converted to thermodynamic temperature units, normalized to unity at $K$-band, and is assumed to be a power-law in antenna temperature with $\beta = -2.14$.
The synchrotron spectral index (relative to $K$-band) is found via
\begin{equation}
\beta_s = \frac{ \log\left[ 0.67\, b_1(\nu) a(\nu) \right] }{ \log(\nu/\nu_{K}) }
\end{equation}
where $a(\nu)$ is the conversion factor from antenna temperature to thermodynamic units.

The coefficients of the model fit to the seven-year data are presented in
Table \ref{tab:template}.  Small changes in the seven-year coefficients compared to previous
values reflect small changes in the absolute calibration and beam profiles. 

For polarization cleaning the maps are degraded to low resolution ($N_\mathrm{side}$ = 16).  The model has the form
\begin{equation}
  [Q(\nu,p), U(\nu,p)]_\mathrm{model} = a_1(\nu) [Q(p), U(p)]_{K} + a_2(\nu) [Q(p), U(p)]_\mathrm{dust}
\end{equation}
The templates used are the \WMAP\ $K$-band polarization for synchrotron ($[Q,U]_{K}$), and a low resolution version of the dust template used above with polarization direction derived from starlight measurements ($[Q,U]_\mathrm{dust}$) and a geometric suppression factor to account for the magnetic field geometry \citep{page/etal:2007}.   The coefficients of the model fit to the seven-year data are in Table \ref{tab:poltemplate}.    For polarization, the template maps are assumed to have a one-to-one correspondence with foreground emission, so the spectral indices for synchrotron and dust are simply the power-law slopes of the coefficients $a_1(\nu)$ and $a_2(\nu)$.  If the dust model is correct then the ratio $a_2/b_3$ gives the polarization fraction; for $W$-band this is $\sim6\%$.

The full-resolution ($N_\mathrm{side} = 512$) foreground-reduced Stokes Q and U maps were produced using the same cleaning coefficients as derived for the low-resolution maps, but with full-resolution templates.  The $K$-band and dust intensity templates can be produced at full resolution from available data, and the starlight polarization map used to determine polarization direction was upgraded to full resolution using nearest-neighbor sampling.  
The templates subtracted from \WMAP\ data are smoothed to $1\degrees$ FWHM, potentially leaving artifacts in the foreground-reduced maps due to small-scale power or beam asymmetries.  In practice, we find no sign of these effects, as discussed in Section~\ref{ssec:template_resids} and Section~\ref{sec:tests}.
All data sets used for templates are available on the LAMBDA Web site.

%%% MCMC fitting %%%%%%%%%%%%%%%%%%%%%%%%%%%%
\subsection{Markov Chain Monte Carlo Fitting\label{sec:mcmc}}
We again perform a pixel-based Markov chain Monte Carlo (MCMC) fitting technique to the five bands of \WMAP\ data.  Our method is similar to that of \cite{eriksen/etal:2007}, but we focus more on Galactic foregrounds rather than CMB. The fit results of the five-year release have been reproduced, with the ``base'' model, which uses three power-law foregrounds, producing virtually the same reduced $\chi^2$ per pixel. 
The MCMC fitting has benefited from further understanding of the zero point of the maps.  We have used the 408 MHz map of \cite{haslam/etal:1981} with a zero-point determined using the same $\csc |b|$ method as for the \WMAP\ data, and investigated the effect on the fit of error in this zero-point.

The MCMC fit is performed on one-degree smoothed maps downgraded to HEALPix $N_\mathrm{side}=64$. A MCMC chain is run for each pixel, where the basic model is
\begin{equation} \label{eq:basemodel}
	T(\nu) = 
	T_{s} \left( \frac{\nu}{\nu_{K}} \right)^{\beta_s(\nu)}
	+ T_{f} \left( \frac{\nu}{\nu_{K}} \right)^{\beta_f} 
	+ a(\nu) T_\mathrm{cmb} + T_{d} \left( \frac{\nu}{\nu_{W}}\right)^{\beta_d}
\end{equation}
for the antenna temperature.  The subscripts $s,f,d$ stand for synchrotron, free-free, and dust emission, $\nu_{K}$ and $\nu_{W}$ are the effective frequencies for $K$- and $W$-bands ($22.5$ and $93.5$ GHz), and $a(\nu)$ accounts for the slight frequency dependence of a $2.725$ K blackbody using the thermodynamic to antenna temperature conversion factors found in \cite{bennett/etal:2003c}.  
The fit always includes polarization data as well, where the model is
\begin{equation}
	Q(\nu) = Q_{s} \left( \frac{\nu}{\nu_{K}} \right)^{\beta_s(\nu)} 
	+ Q_{d} \left( \frac{\nu}{\nu_{W}}\right)^{\beta_d} 
	+ a(\nu) Q_\mathrm{cmb}
\end{equation}
\begin{equation}
	U(\nu) = U_{s} \left( \frac{\nu}{\nu_{K}} \right)^{\beta_s(\nu)} 
	+ U_{d} \left( \frac{\nu}{\nu_{W}}\right)^{\beta_d} 
	+ a(\nu) U_\mathrm{cmb}
\end{equation}
for Stokes Q and U parameters.  Thus there are a total of 15 pieces of data for each pixel ($T$, $Q$, and $U$ for five bands).

As for the five-year release, the noise for each pixel at $N_\mathrm{side}=64$ is computed from maps of $N_\mathrm{obs}$ at $N_\mathrm{side}=512$.  To account for the smoothing process, the noise is then rescaled by a factor calculated from simulated noise maps for each frequency band.  The MCMC fit treats pixels as independent, and does not use pixel-pixel covariance, which leads to small correlations in $\chi^2$ between neighboring pixels.  This has negligible effect on results as long as goodness-of-fit is averaged over large enough regions.

We fit three categories of models.  All use $K$-band as a template for the polarization angle of synchrotron and dust emission, so $U_s$ and $U_d$ are not independent parameters, identical to the previous analysis.  All models also fix the free-free spectral index to $\beta_f = -2.16$, a slight change from $\beta_f=-2.14$ used in the previous analysis.  This change was motivated as an attempt to better match the effective spectral index at $Q$- and $V$-bands, due to their use in cosmological analysis, but was not found to make a difference to the fits.

The ``base'' model uses three power-law foregrounds, where the synchrotron spectral index $\beta_s(\nu)$ is taken to be independent of frequency but may vary spatially, and the dust spectral index $\beta_d$ is allowed to vary spatially.  We assume the same spectral indices for polarized synchrotron and dust emission as for total intensity emission.  This model has a total of 10 free parameters per pixel: $T_s$, $T_f$, $T_d$, $T_\mathrm{cmb}$, $\beta_s$, $\beta_d$, $Q_s$, $Q_d$, $Q_\mathrm{cmb}$, and $U_\mathrm{cmb}$.

A steepening synchrotron model uses the same three foregrounds but allows for a steepening of the synchrotron spectral index by adding a new parameter $c_s$, defined by
\begin{equation}
	\beta_s(\nu) = \left\{ \begin{array}{cr}
		\beta_s & \nu < \nu_{K} \\ 
		\beta_s + c_s \ln\left(\frac{\nu}{\nu_{K}}\right) & \nu > \nu_{K}
	\end{array}\right..
\end{equation}
For the steepening model the dust spectral index is fixed\footnote{The precise choice of dust index here and for the spinning dust model does not make much difference; when allowed to vary it is poorly constrained by the MCMC fits and uncorrelated with the synchrotron or free-free components \citep{gold/etal:2009}.} to $\beta_d=+2.0$.  Therefore this model also has 10 free parameters per pixel.

For models with a spinning dust component, another term is added to equation~\ref{eq:basemodel}
\begin{equation}
	\label{eq:sdfunc}
	 T_{sd}(\nu) = A_{sd} \frac{(\nu/\nu_{sd})^{\beta_d+1}}{\exp(\nu/\nu_{sd})-1}.
\end{equation}
The spinning dust component is assumed to have negligible polarization, as theoretical expectations for the polarization fraction are low compared to synchrotron radiation \citep{lazarian/draine:2000}, and the polarization data thus far show no evidence that such a component is necessary (see Section~\ref{ssec:haze}).  The spinning dust amplitude $A_{sd}$ was allowed to vary spatially as a new parameter.  Both $\beta_s$ and $\beta_d$ were fixed to $-3.0$ and $+2.0$, respectively, to avoid degeneracies from having too many parameters in the fit. Allowing $\nu_{sd}$ to spatially vary was not found to result in any improvement of the fit, but fits were performed with different global values of $\nu_{sd}$ to find the best overall value.  Thus with fixing of the spectral indices, this model has nine free parameters per pixel.

MCMC fits for the seven-year release were performed with the addition of the 408 MHz data compiled by \cite{haslam/etal:1981}.  The error on the zero point for this data was estimated in that work to be $\pm 3$ K, with an overall calibration error of $10\%$. \cite{lawson/etal:1987} use a comparison with 404 MHz data to find a uniform (presumably extragalactic) component with a brightness of $5.9$ K.  As the MCMC method treats all input maps equally, for consistency we estimate and subtract off a nominal zero point offset of 7.4 K, as determined by the same $\csc|b|$ method we use for the \WMAP\ sky maps.  However,  the 408 MHz data resembles a $\csc|b|$ behavior much less than the \WMAP\ data, due to the increased relative prominence of large-scale features such as the Northern Polar Spur.  Therefore we attempted the $\csc|b|$ fitting procedure on different hemispheres and with different cuts, and estimate the uncertainty in procedure to be $\pm 4$ K.  MCMC fits were run for each model with zero points of 3.4 K and 11.4 K in addition to the nominal value, and the effect of these on foregrounds is discussed in Section~\ref{ssec:sd}.  A full set of maps and MCMC variance estimates for the three models is available on the LAMBDA Web site.

%%% EXT data
\section{Comparison with ARCADE 2\label{sec:arcade}}
The ARCADE collaboration has made available absolute temperature measurements of Galactic emission for part of the sky \citep{kogut/etal:2010}.  
ARCADE observations do not cover the full sky and the instrument's beam is significantly larger than \WMAP's.  Therefore we limit our comparison to two regions where ARCADE's scan crosses the Galactic plane and observes the brightest emission, the first at Galactic longitude of $34\degrees$ and the second at 
$93\degrees$.

\begin{figure}
\epsscale{0.75}
%\plotone{sdbypix2.pdf}
\plotone{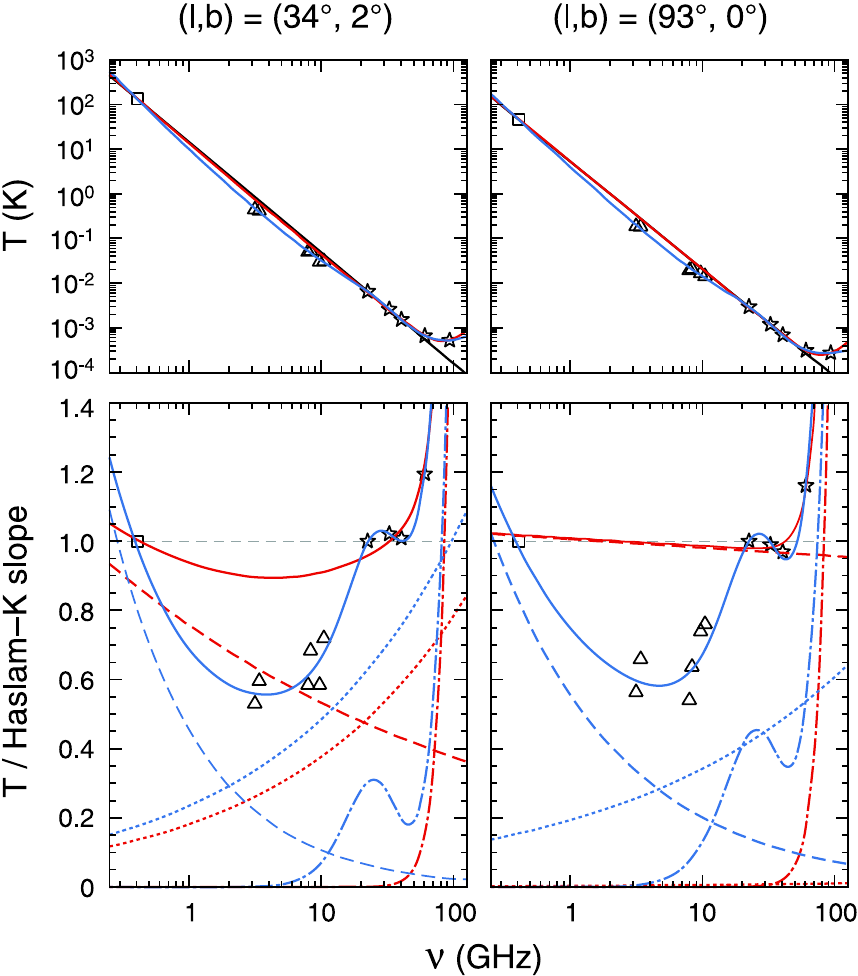}
\caption{\label{fig:arcade}
Galactic emission from two regions in the Galactic plane.  ARCADE (triangles), \WMAP\ (stars), and 408 MHz data (square) are all shown, smoothed to a common resolution.  Upper panels show antenna temperature (absent a monopole component).  The black line is a power-law connecting 408 MHz to 22 GHz ($\beta=-2.48$ for the left panel, $\beta=-2.41$ for the right panel), which is divided out in the bottom panels to better show deviations from power-law behavior.  Red lines show the result of a fit to the data using three power law components for foregrounds (representing synchrotron, free-free, and dust).  Blue lines show the fit resulting when an extra component representing spinning dust is added.  Solid lines show the total flux, with individual components shown by dashed lines (synchrotron), dotted lines (free-free), and dot-dashed lines (dust plus spinning dust).  Errors in the data are dominated by systematics and highly correlated between data points, but are estimated to be $5-15\%$, depending on experiment.
}
\end{figure}

Figure \ref{fig:arcade} shows the Galactic spectrum for these two regions.  \WMAP\ data have been smoothed to $11.6\degrees$, to match the ARCADE resolution.  
The ARCADE maps have had the CMB monopole removed, and the WMAP maps have had CMB anisotropies removed using the ILC map (though this has little effect).  
The ARCADE data have \emph{not} had any extragalactic component (as found by \citealt{fixsen/etal:2010}) removed.  
Instead, all maps have been treated as equally as possible, removing a zero-point by fitting a $\csc|b|$ model to the available data and subtracting the constant term.  

The uncertainty in this zero-point subtraction is largest for ARCADE due to the limited sky coverage of the experiment.  We tested the zero-point subtraction by fitting to several partial-sky subsets of the full-sky \WMAP\ maps, and find that the variations imply an uncertainty in the ARCADE points of up to 15\% of the CMB-subtracted flux.
Also included is the 408 MHz map as a reference point at low frequency.  As discussed in the previous section, the $\csc|b|$ model performs most poorly for this map, with uncertainties of $\pm 4$ K.  However, in these two regions the emission is bright enough that this is still less than $10\%$ of the total emission.

Two fits were applied to the data in each region.  The first used three power-law foregrounds: synchrotron, free-free, and dust, where the spectral indices for synchrotron and dust were left free for the fit.  The second fit added a spinning dust component using the functional form of Equation~\ref{eq:sdfunc}, with the amplitude and $\nu_{sd}$ as free parameters.  Because the maps are highly smoothed, errors are dominated by systematic issues and difficult to characterize.  We chose to use $2\%$ fractional error for \WMAP\ and $5\%$ fractional error for other observations when performing the fit.
Using larger errors does not remove the sharp difference in $\chi^2$ between the two models unless the errors are taken to be larger than $50\%$.  
For the fit without spinning dust, the ARCADE data were not used in the fit as they were found  to be incompatible with such a model.

The resulting fits are shown in Figure~\ref{fig:arcade}, with the spinning dust fit in blue and the power-law-only fit in red.  The top panels show the data and fits in absolute temperature units after monopole subtraction.  The bottom panels show the same data and fits, but where all temperatures have had the 0.408--22 GHz slope divided out, to facilitate comparison with Figure~9 of \cite{kogut/etal:2010}.  The ARCADE data show a clear deficit over the 3--10 GHz range, which cannot be explained with power-law foregrounds alone; a fit including a spinning dust component is much more consistent.  Dotted, dashed, and dash-dotted lines in the figure show the contribution of each individual component to the total, with thermal dust and spinning dust shown together.  In the spinning dust model, synchrotron emission is weak in the \WMAP\ bands, where free-free is the dominant emission process.  At $93\degrees$ longitude the spinning dust emission is approximately as bright as the free-free emission at 23 GHz, and at $34\degrees$ longitude it is several times fainter at all frequencies.

\section{Foreground Results\label{sec:results}}

\subsection{Residuals in Template-Cleaned Maps\label{ssec:template_resids}}

\begin{figure}
\epsscale{1.0}
%\plotone{fig_q-w.pdf}
\plotone{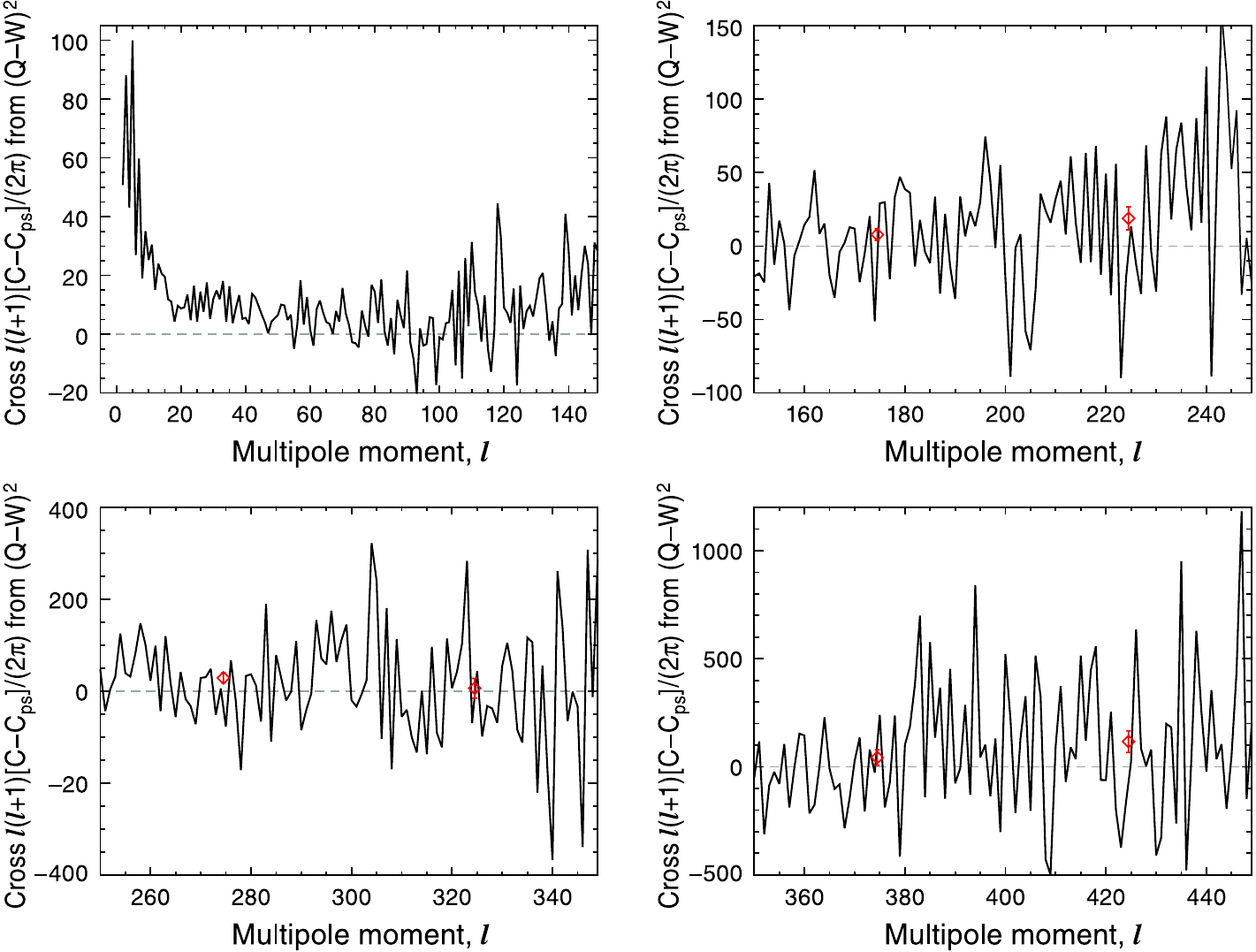}
\caption{\label{fig:qwdiffcleanps}
Power spectrum of the difference between foreground-reduced maps.  $Q$-band minus $W$-band is shown here, with a point source contribution subtracted off.  Note the changing scale between panels.  Red points with error-bars are averages over bins with $\Delta \ell = 50$.  Deviations from zero are below 100 $\mu\textrm{K}^2$ outside the KQ85y7 mask, and the upper bound to foreground contamination in the foreground-reduced maps is $15$ $\mu\textrm{K}$.
}
\end{figure}

As a test of the template-based foreground subtraction process, power spectra of difference maps were made.  Figure \ref{fig:qwdiffcleanps} shows the power spectrum of the difference between the foreground-reduced $Q$-band and $W$-band maps, with the point source contribution to the power spectrum subtracted off.  Averaging over bins of $\Delta \ell = 50$, no bin with more than 120 $\mu$K${}^2$ of power is seen, with an upper limit of $\sim 220$ $\mu$K${}^2$ in power (15 $\mu\textrm{K}$ in amplitude), and the results are consistent with zero within the expected error.  For comparison, CMB power in the range $30 < \ell < 500$ is $1000$ $\mu$K${}^2$ or more \citep{larson/etal:prep}.
Differences between foreground-reduced $V$-band and $W$-band were also computed, and the power in that case was even smaller.

%%% FG powerspec %%%%%%%%%%%%%%%%%%%%%%%%%%%
\subsection{Polarization Power Spectra of Synchrotron and Dust\label{ssec:fgpowerspec}}
While Galactic foregrounds are not fully described by a two-point function (i.e. an angular power spectrum), due to the importance of the CMB it is often useful to examine the power spectrum of foregrounds.  Specifically, the relevant quantity to calculate is the contribution of foreground emission to the angular power spectrum in a particular patch of sky of interest for CMB analysis.

\begin{figure}
\epsscale{0.8}
%\plotone{mcmcfg_cl.pdf}
\plotone{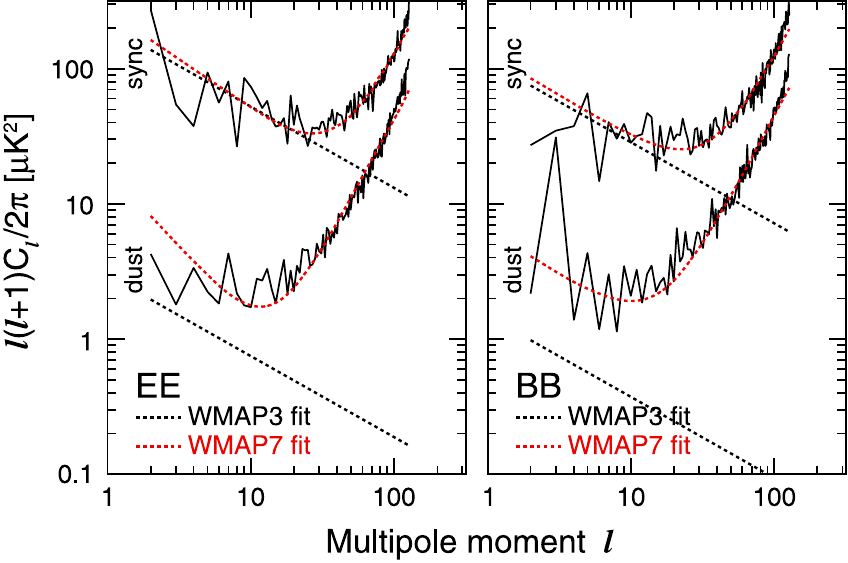}
\caption{\label{fig:fgpowerspec}
Power spectra of polarized foreground components as determined by the MCMC model.  On the left are $C_\ell^{EE}$ and on the right are $C_\ell^{BB}$; for foregrounds these should be of comparable magnitude. The black dotted lines are the foreground fit to raw three-year WMAP data from \cite{page/etal:2007}, and the red dotted lines are the combined foreground and noise fit to MCMC maps from this work, with coefficients given in Table~\ref{tab:fgpowerspec}.  Synchrotron results are in good agreement with the previous analysis.  The seven-year dust results spectra appear to have a higher amplitude, but the signal-to-noise for $\ell \ge 10$ is 2.8 or less for dust.
}
\end{figure}

A general trend of $\ell (\ell+1) C_\ell \sim \ell^{-0.6}$ was found from examination of raw polarization data outside the P06 mask \citep{page/etal:2007}, as the result of a combined fit to \WMAP\ data in both multipole and frequency space.
With the MCMC fitting procedure it is possible to separate polarized synchrotron from dust and examine the two components individually. The results, shown in Figure~\ref{fig:fgpowerspec}, show behavior largely consistent with the previous analysis.

In detail, MCMC maps from the ``base'' model including Haslam data were used.  Power spectra from the spinning dust model MCMC maps were also inspected and found to be nearly identical at large scales. A union of the polarization analysis mask and the mask of pixels flagged by the MCMC was applied, and the $C_\ell^{EE}$ and $C_\ell^{BB}$ spectra were computed for both synchrotron and dust.  As the MCMC process uses one-degree smoothed maps, an appropriate correction for the beam window function was applied. Each power spectrum was then fit with a model consisting of a power-law plus a pixel noise term
\begin{equation}
	\ell(\ell+1)C_\ell^{XX}/2\pi = {\cal B}_c \ell^m + \ell(\ell+1)N^2,
\end{equation}
where ${\cal B}_c$ is the amplitude for foreground component $c$, $m$ is the power-law index, and $N$ the noise amplitude.  

Values for the fit parameters and an estimate of their errors can be found in Table~\ref{tab:fgpowerspec}.  Because the power spectra are taken from highly processed maps, detailed error propagation is difficult.  We used the diagonal portion of the published $C_\ell$ Fisher errors plus cosmic variance to perform the fit; covariance between multipoles will cause the true errors to be somewhat larger.  If all foreground power spectra are assumed to have the same power-law behavior, then the weighted mean $m = -0.67\pm0.24$.

%%% free-free? %%%%%%%%%%%%%%%%%%%%%%%%%%%%
\subsection{Free-free Emission}
That the ratio of radio brightness to H$\alpha$ intensity from the MEM fits is consistently lower than the expected value has long been of concern.  The MCMC fits offer some insight, though unfortunately do not  resolve  the difference.
The most important difference between the MEM and the MCMC fits in this case is that the MEM uses the H$\alpha$ template as a prior in low signal-to-noise regions, while the MCMC fit does not.  The result is that in regions of low signal, the degeneracy between synchrotron and free-free causes the MCMC uncertainty in free-free brightness to be large enough to accommodate a large range of possible radio to H$\alpha$ ratios.  Therefore it becomes necessary to exclude low signal-to-noise regions when calculating the ratio from the MCMC maps.

Due to the uncertainty in the reddening correction to the H$\alpha$ map itself, it is also customary to exclude regions where the H$\alpha$ optical depth due to reddening is greater than some value.  This unfortunately excludes regions that would otherwise have \emph{high} signal-to-noise.
These two cuts together exclude much of the MCMC maps as unsuitable for analysis.  The remaining portion of sky contains bright, mostly discontiguous free-free regions which are also low on dust (and therefore H$\alpha$ extinction).  The largest of these is a region around Gum nebula.  

Starting with the free-free maps made from the MCMC process, we define a signal-to-noise ratio (SNR) map as the free-free amplitude divided by the square root of the MCMC variance.  We then keep only pixels with $\textrm{SNR}>10$, $\tau < 1$, and no MCMC error flags.  The pixels that remain are largely concentrated in three regions, the Gum nebula, the Ophiuchus complex, and the Orion/Eridanus bubble.  The Gum region contains nearly half of the pixels surviving the cut, so for simplicity we restrict our attention to this region, defining it to be any pixel within $30\degrees$ of ($260\degrees, 0\degrees$) in Galactic coordinates.  Summing all free-free emission in this region and dividing by the total H$\alpha$ intensity in this region, we estimate that the ratio of radio brightness to H$\alpha$ intensity $h_\mathrm{ff}$ is $9.3 \pm 3.2$ $\mu\textrm{K R}^{-1}$ at $K$-band for the spinning dust fit, with similar values for the other MCMC models.  The uncertainty comes from the variance of the ratio from pixel to pixel; increasing the signal-to-noise threshold  decreases the uncertainty somewhat but does not significantly affect the central value.  While the central value is consistent with the prediction of $11.4$ $\mu\textrm{K R}^{-1}$ within this uncertainty, it is also compatible with a reduced electron temperature of 5500 K, an overestimation of the reddening correction by $\Delta\tau = 0.3$, or some combination of the two.

%%% spinning dust %%%%%%%%%%%%%%%%%%%%%%%%%
\subsection{Spinning Dust Emission\label{ssec:sd}}

We find that in order to best fit the 408 MHz data, the spinning dust fit from the five-year MCMC process needs to have its peak frequency adjusted downward by $14\%$ from $\nu_{sd}=4.9$ GHz to $\nu_{sd} = 4.2$ GHz, nearly independent of the offset used for the map.  For this value, the frequency at which the \emph{flux} from the spinning dust component alone peaks is $21$ GHz.  We have not found any improvement in the fit from including `warm' spinning dust with a peak near $40$ GHz, as found by \citealt{dobler/finkbeiner:2008b}.  The 408 MHz data also introduces some tension, such that the spinning dust model no longer is such a large improvement inside the Galactic plane; in this region $\chi^2_\nu = 1.80$ for the spinning dust model, compared to $\chi^2_\nu = 2.61$ for the ``base'' fit, for $8.7$ effective number of degrees of freedom (see \citealt{kunz/trotta/parkinson:2006, gold/etal:2009} for  detailed description of effective d.o.f.).
These values are for the fitted offset of $7.4$ K for the 408 MHz map; using a larger offset value of $11.4$ K provides slightly better fits ($\Delta \chi^2_\nu = 0.008$) for the spinning dust models, while a smaller offset value of $3.4$K provides slightly better fits ($\Delta \chi^2_\nu = 0.074$) for models without spinning dust.

The steepening synchrotron model fits the combination of \WMAP\ and 408 MHz data nearly equally as well as the spinning dust model, with $\chi_\nu^2 = 1.81$ in the Galactic plane.  The amplitude of  $c_s$ is large in the Galactic plane, implying a change of spectral index greater than one per $e$-fold increase in frequency.  This is a sharper change than models of synchrotron steepening predict from aging effects, and so the physical motivation for the model is unclear.

The ARCADE data are not directly comparable to the MCMC fits, due to their greatly different beam and sky coverage.  The spinning dust component of the fits for the two regions in the Galactic plane, however, does peak in flux at $22$ GHz, consistent with the location of the MCMC peak.  The relative amplitude is more difficult to ascertain. For ARCADE, spinning dust is $29\%$ (at $l=33.8$) or $43\%$ (at $l=93$) of the total flux at $22$ GHz, but with the large beam it is impossible to say whether the spinning dust component is relatively diffuse or localized on the Galactic plane.  For the MCMC fits to \WMAP\ data, the mean spinning dust fraction is considerably lower, at $18\%$ inside the KQ85y7 mask, which suggests that spinning dust may be patchy.  Outside of the KQ85y7 mask, the MCMC fits show a mean level of spinning dust consistent with zero within the uncertainty of the fit.

%%% Haze %%%%%%%%%%%%%%%%%%%%%%%%%%%%%%
\subsection{The Haze\label{ssec:haze}}
In its low frequency bands, \WMAP\ observes an excess of emission above what was predicted by scaling the 408 MHz to higher frequencies using the expected spectral index for synchrotron emission.  Determining the exact nature of this emission has proven difficult; \WMAP\ has generally treated it as a hard (flatter spectrum) synchrotron component without attempting to explain the origin of such a component.  Other suggestions have involved combinations of different types of spinning dust \citep{finkbeiner:2004,dobler/finkbeiner:2008b}, though there is typically still a residual ``haze'' even after those components are fit out \citep{dobler/finkbeiner:2008a}.

It has been argued that this remainder low-frequency emission has an ellipsoidal shape and is consistent with hard synchrotron emission, possibly from dark matter annihilation in the core of the Galaxy \citep{hooper/finkbeiner/dobler:2007}.  There has been tentative detection of a haze in gamma-rays using preliminary data from the Fermi telescope \citep{dobler/etal:2009}. 

\begin{figure}
\epsscale{0.8}
%\plotone{polfit_comparison_v2.pdf}
\plotone{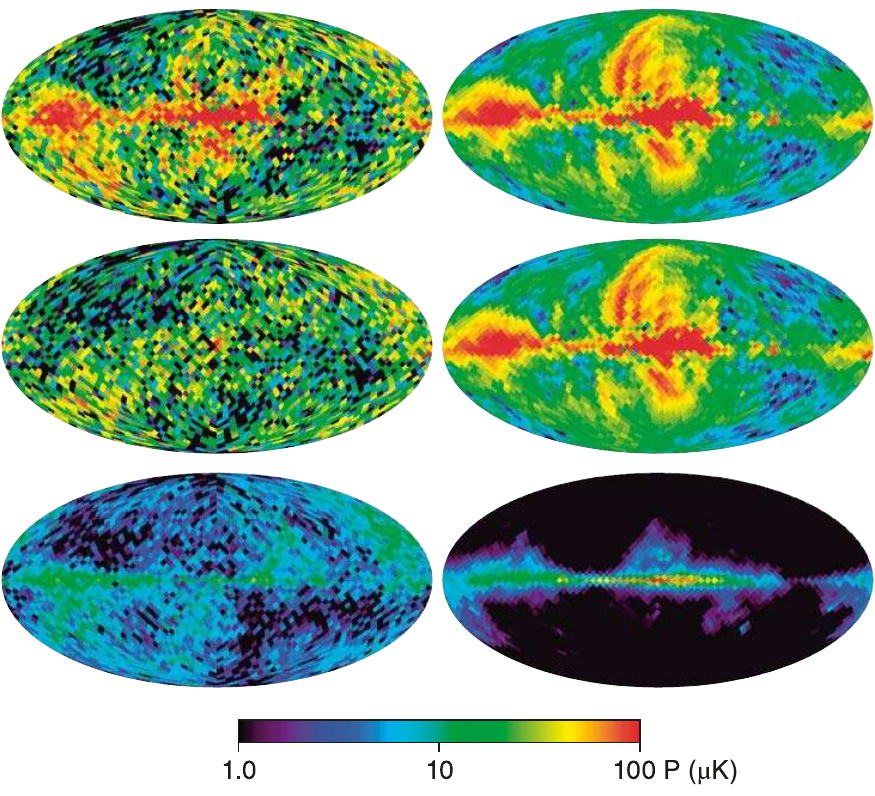}
\caption{\label{fig:polhaze}
Comparison of the templates used for polarization cleaning to a low-resolution ($N_\mathrm{side} = 16$) MCMC fit to polarization data using a three-component model with fixed spectral indexes to search for any hard synchrotron component.  
The left column shows the results of the MCMC fit to polarization data using three components: soft synchrotron ($\beta = -3.1$) at top, hard synchrotron ($\beta=-2.39$) at middle, and dust ($\beta=+2.0$) at bottom.  For comparison, the right column shows the templates used for polarization cleaning: synchrotron at top and middle, and dust at bottom.  All plots are of polarization intensity $P=\sqrt{Q^2+U^2}$, with a logarithmic scale from 1 to 100 $\mu$K.  Synchrotron intensity is measured at a reference frequency of 23 GHz, and dust intensity at 94 GHz.  The MCMC maps are noisy, and have been corrected for a noise bias in $P$ caused by noise in $Q$ and $U$.  Excess noise in the plane of the ecliptic due to the scan pattern is also clearly visible in the MCMC fits.  
Given the noise level, hard synchrotron emission does not appear to be significant.
}
\end{figure}

Interpretation of polarization information toward the center of the Galaxy is difficult, as depolarization through line-of-sight changes in the orientation of the magnetic field can affect the signal significantly.  Nonetheless, we search for a hard component in the polarization data using a simplified version of the low-resolution MCMC fit of \cite{dunkley/etal:2009}, shown in Figure~\ref{fig:polhaze}.  We do not detect any significant change of synchrotron spectral index as a function of Galactocentric distance.

This special fit was done at HEALPix $N_\mathrm{side}=16$ using \emph{only} \WMAP\ polarization data, so as to be insensitive to any uncertainties regarding the presence or absence of spinning dust.  The fit attempts to model the sky as a sum of three power-law foregrounds: a soft synchrotron component with $\beta=-3.1$, a hard synchrotron component with $\beta=-2.39$, and a dust component with $\beta=+2.0$.  These power-law indices were those suggested by the work of \cite{dobler/finkbeiner:2008a}.  

The results of the fit are shown in Figure~\ref{fig:polhaze}.  Residuals after the fit are small compared to the noise, and over all bands the mean reduced $\chi^2$ per pixel is $1.1$.
For comparison, the synchrotron and dust templates used for polarization cleaning are shown in the right column of the figure.  The MCMC result for the soft synchrotron template appears to be essentially a noisy version of the synchrotron template, indicating that $K$-band indeed is a good proxy for polarized synchrotron emission.  For dust, the MCMC and template results differ somewhat.
The MCMC hard synchrotron results show no spatial structure beyond \WMAP's noise pattern, and are consistent with the level of noise bias expected in a map of $P=\sqrt{Q^2+U^2}$.

\begin{figure}
\epsscale{0.7}
%\plotone{gc.pdf}
\plotone{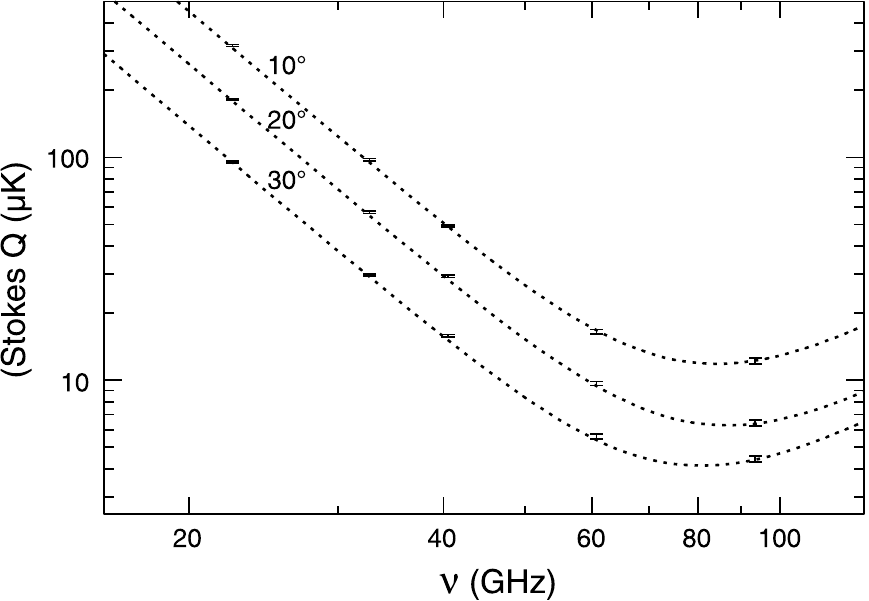}
\caption{\label{fig:gcspec}
Frequency spectrum of polarized emission around the Galactic center.  Average antenna temperature of Stokes Q is shown for three oval regions defined by $\sqrt{l^2+(2b)^2} < 10\degrees,20\degrees,30\degrees$, where $l$ and $b$ are Galactic longitude and latitude.  Stokes U is negligible at all frequencies except $W$-band.  Errorbars indicate statistical uncertainty from the diagonal part of the pixel-pixel noise matrix.  Dotted lines show the sum of a synchrotron component with $\beta = -3.2$ and a dust component with $\beta = +2.0$; in all cases this two-component model is sufficient to explain the observations.
}
\end{figure}

Figure~\ref{fig:gcspec} shows the frequency spectrum of polarized emission for elliptical regions around the Galactic center.  In these regions the polarization direction is nearly vertical, and so the Stokes U parameter is negligible for bands $K$ through $V$ and small for $W$-band.  The spectra for three different regions are shown, sized $10\degrees\times5\degrees$, $20\degrees\times10\degrees$, and $30\degrees\times15\degrees$.  We find no evidence for emission other than soft synchrotron ($\beta=-3.2$) and dust ($\beta = +2.0$), in particular, no ``haze'' component appears to be necessary for polarization.  

%%% Systematics %%%%%%%%%%%%%%%%%%%%%%%%%%%%
\section{Foreground Systematics and Tests\label{sec:tests}}
\subsection{Pipeline Simulation}
A full simulation of the \WMAP\ instrument was used to test the MCMC and template cleaning methods, in order to investigate the interaction of systematics in both time-domain data and sky maps.  
Starting with a set of synthetic sky inputs for the CMB and foregrounds (described below), the scanning of the instrument was applied to the inputs to produce a timestream of data, which was then put through the same entire calibration and map-making pipeline as used for real data.

A random CMB realization was created, starting from the publicly available best-fit cosmological parameters of a $\Lambda$CDM model to the combination of five-year \WMAP\ data with supernovae and baryon acoustic oscillations. The \textsc{CAMB} software package \citep{lewis/challinor/lasenby:2000} was used to generate a model power spectrum and then \textsc{synfast} \citep{gorski/etal:2005} was used to generate the random sky realization.

Several foregrounds were then added, using high resolution templates.  A synchrotron intensity template was constructed from the 408 MHz data of \cite{haslam/etal:1981}, and scaled to higher frequencies with a spectral index with both spatial variations and steepening, in order to test the effects of fitting a simpler model to complicated synchrotron spectral features.  A free-free template was made from an extinction-corrected version of the H$\alpha$ map of \cite{finkbeiner:2003}, with a few bright high-latitude sources removed, and assuming a spectral index of $\beta = -2.15$.   The dust template is the 94 GHz prediction of model 8 of \cite{finkbeiner/davis/schlegel:1999}, scaled to other \WMAP\ frequencies with a spectral index of $\beta=2.0$. 

Once the simulation inputs were generated, they were passed through a simulation of \WMAP's scan strategy, including such effects as thermal gains and baselines in the time-ordered data, loss imbalance and bandpass mismatches, and detector noise with a $1/f$ component.  This simulated time-ordered data was then processed and analyzed in exactly the same way as real observations.

\begin{figure}
\epsscale{0.8}
%\plotone{simdiff_band0_v2.pdf}
\plotone{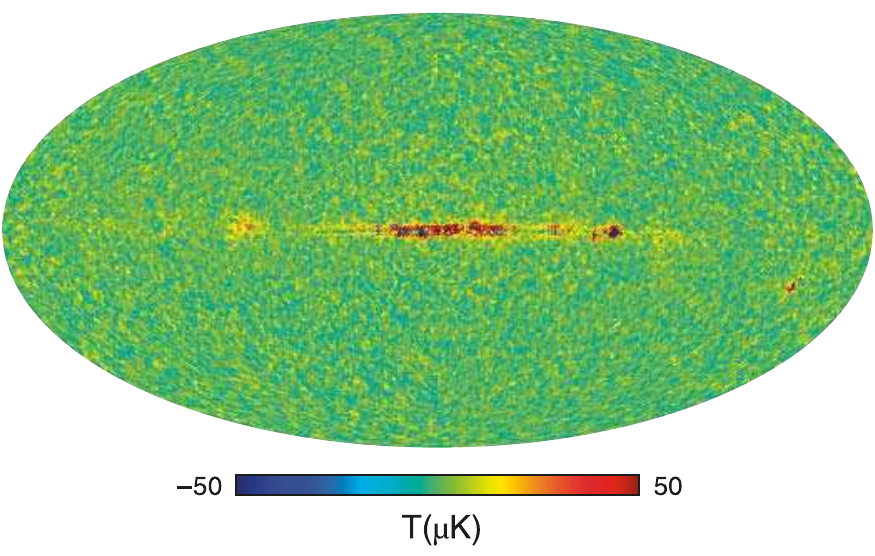}
\caption{\label{fig:simp4a}
Comparison between a simulated input sky and the resulting maps after scanning and map-making.  $K$-band is shown; differences in other bands are at least four times smaller.  The only visible structure, along the Galactic plane, is entirely consistent with residuals from gain reconstruction within the quoted uncertainties (0.2\%).
}
\end{figure}

\begin{figure}
\epsscale{0.9}
%\plotone{simdiffmc.pdf}
\plotone{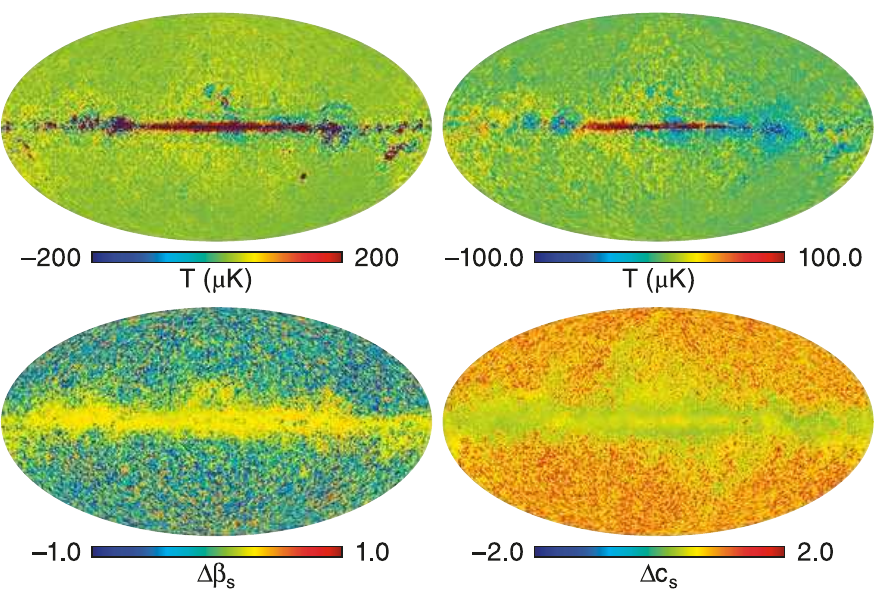}
\caption{\label{fig:simp4a_mcmc}
Comparison between the input foreground spatial and spectral behavior and that recovered by the MCMC fit. Upper left: difference between MCMC result and input $T_s+T_f$.  Upper right: difference between MCMC result and input $T_d$.  Lower left: difference between MCMC result and input $\beta_s$.  Lower right: difference between MCMC result and input $c_s$.  The main feature is that the simulated synchrotron model contained more steepening in the synchrotron spectrum than the model allowed for, which then biases the recovered $\beta_s$ by $0.2$ in high signal-to-noise regions.  The apparent bias off the Galactic plane only occurs where the signal-to-noise is low and the parameter error is larger than the bias.
}
\end{figure}

Figure~\ref{fig:simp4a} shows a comparison  between the ``true'' simulated input sky maps and the output maps after the map-making process.  These are used to test the template cleaning method, as the simulated input foregrounds are generated with structure on scales smaller than the templates used for cleaning.  However, no effects due to residual foreground contamination are seen; the cosmological parameters used as input are recovered.
Figure~\ref{fig:simp4a_mcmc} then compares the results of the MCMC foreground fit to the input foreground behavior.
The largest difference found between the input and output maps from the simulation is in the Galactic plane.  This difference is a fraction of a percent of the total intensity, and is entirely consistent with the expected uncertainty in the gain reconstruction.

The MCMC reconstructs the foregrounds to within the MCMC error, which includes large covariance between synchrotron and free-free brightness.  The most important systematic deviation was in the reconstructed synchrotron spectral shape, parameterized with $\beta_s$ and $c_s$. 
This is largely because the simulated model spectrum was more complicated than a power-law with constant steepening.  This resulted in a  bias in the recovered $\beta_s$ bias of approximately $+0.2$ in the Galactic plane.  This bias was still within the MCMC errors.

\subsection{Testing Beam Systematics with Six-month Maps}
Over the course of a full year, the \WMAP\ satellite's scan pattern is such that most points on the sky are observed with a nearly uniform distribution of orientations.  The distribution is most symmetric at the ecliptic poles, and least symmetric on the ecliptic plane.  Fortuitously, the Galactic center lies near the plane of the ecliptic, with a large angle between the planes of the Galaxy and the ecliptic.  The result is that the year can be divided into halves, where \WMAP's scanning direction when observing the inner Galactic plane is rotated $180\degrees$ between the two halves.

\begin{figure}
\epsscale{0.5}
%\plottwo{sixmodiff_beam1.pdf}{sixmodiff_beam2.pdf}\\
%\plotone{sixmodiff_maps.pdf}
%\plottwo{f12a.pdf}{f12b.pdf}\\
%\plotone{f12c.pdf}
\plotone{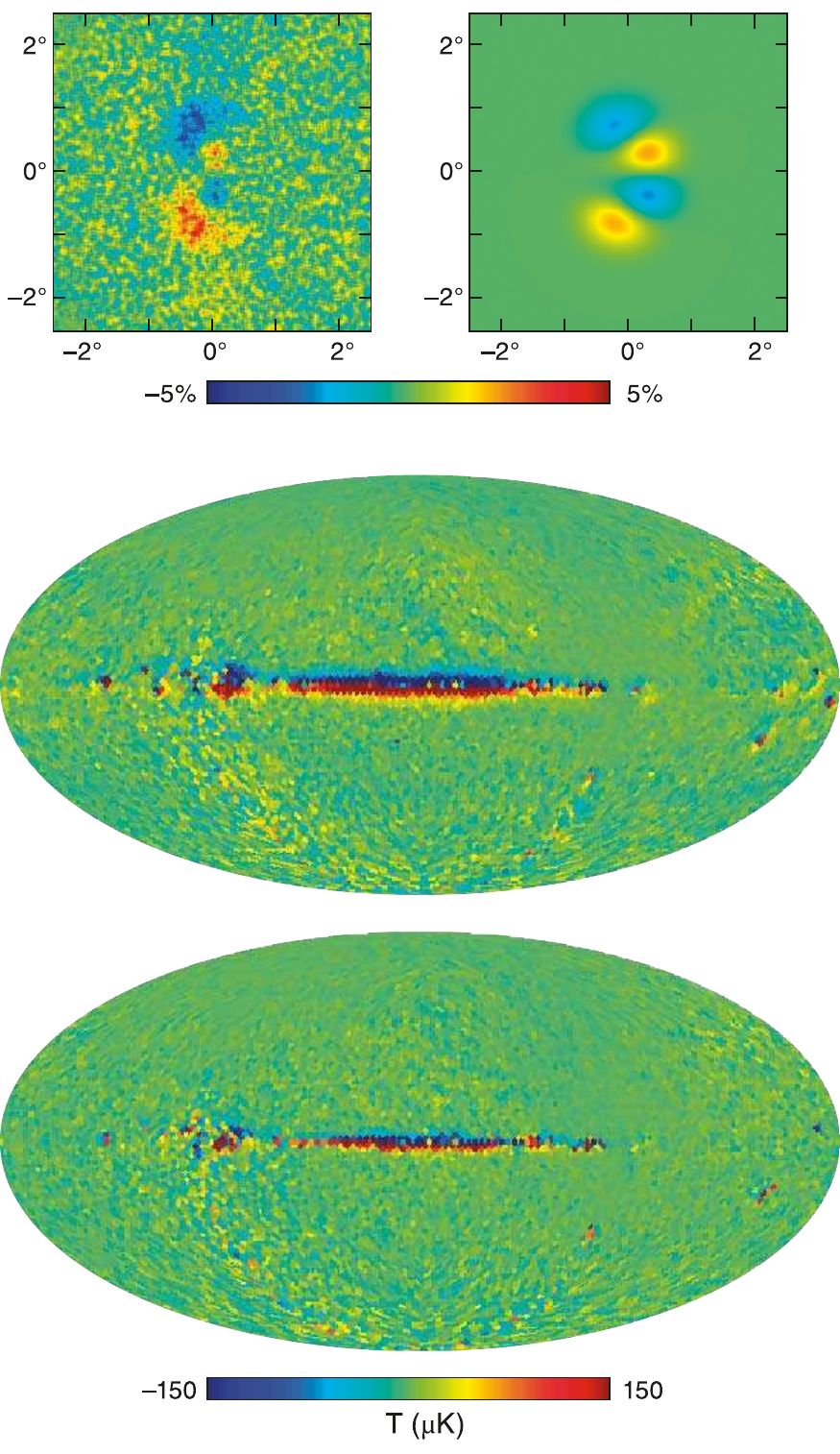}
\caption{\label{fig:sixmodiff_beam+map}
Flight data and a simple model for differences between maps made with six months of data and those made with a full year.  Top left: difference between observed $K$-band beam and $180\degrees$ rotated $K$-band beam (scale is $\pm 5\%$).  Top right: difference between a model beam consisting of a sum of Gaussians and its $180\degrees$ rotation.  Middle: observed difference map between six months and a full year for $K$-band.  Bottom: simulated difference map created using the beam of the upper right panel.  
While this simple beam model does not completely resemble the observed beam, it qualitatively reproduces the effects observed in the maps.
}
\end{figure}

This means that maps made from such six-month segments of data are sensitive to beam asymmetries, particularly those where the beam is not equal to itself rotated $180\degrees$.  This effect is largest in $K$-band.  Figure~\ref{fig:sixmodiff_beam+map} shows the measured difference of the beam between the six-month intervals, a simple beam model which recreates the effect, and sky maps of the residuals between six-month sky maps and a full year of observation.

We used this effect to investigate the sensitivity of foreground fitting to beam systematics.  For the first five years of data, each year was divided into six months of one scan direction relative to the Galactic center, and six months where the scan direction was reversed.  These were then stacked to produce two sets of five-year maps, where the scan directions along the ecliptic have the greatest relative asymmetry.  The MCMC foreground fitting was then run for both sets of maps.

\begin{figure}
\epsscale{0.9}
%\plotone{sixmomc.pdf}
\plotone{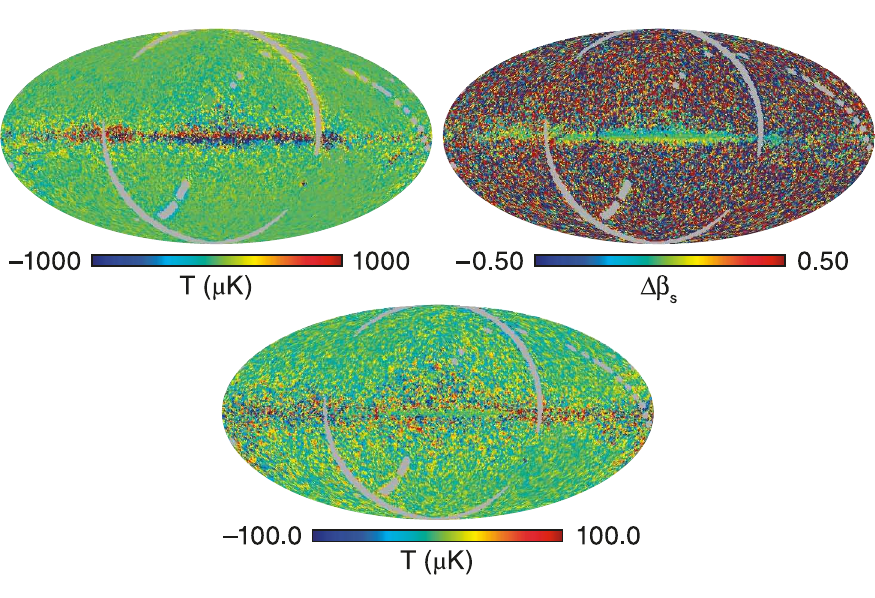}
\caption{\label{fig:sixmodiff_mcmc}
Effect of beam anisotropy on the MCMC foreground fits, using stacks of six-month maps.  Pixels near the boundary of the six-month scans are masked (gray) due to poor coverage.  Top left: difference in MCMC synchrotron temperature.  As the combination of synchrotron and free-free is largely constrained to match $K$-band, the free-free difference is nearly the opposite of this map.  Top right: difference in MCMC synchrotron spectral index.  Away from the Galactic plane this map is mostly noise, but a slight gradient with $\Delta\beta_s = \pm0.1$ is visible near the plane.  Bottom: difference in MCMC CMB temperature. Most of the variation is noise in the MCMC dust model, rather than due differences between the six-month maps.
}
\end{figure}

The result is shown in Figure~\ref{fig:sixmodiff_mcmc}.  Since the largest beam difference is in $K$-band, low frequency foregrounds are most strongly affected.  The spectral index inferred for synchrotron shows a small gradient across the Galactic plane, with amplitude $\sim \pm 0.1$ for $|b| < 5\degrees$.  The effect on the CMB is limited; variance outside the KQ85y7 mask is less than $480$ $\mu\textrm{K}^2$ (an order of magnitude smaller than intrinsic variance of the CMB), and most of this is from MCMC variations in the dust model.  We also emphasize that the six-month intervals were chosen to maximize this asymmetry, which is not seen when full years of data are used to make maps.

%%% Point Sources %%%%%%%%%%%%%%%%%%%%%%%%%%

\section{Point Source Catalogs\label{sec:pointsources}}

As for the five-year analysis, two separate methods have been used for identification of
point sources from skymap data and two separate point source tables have been produced.
The first method has been used in all \WMAP\ data releases and is largely
unchanged from the five-year analysis \citep{wright/etal:2009}.  The seven-year signal-to-noise
ratio map in each wavelength band is filtered in harmonic space by $b_l/(b^2_lC^{\rm cmb}_l
 + C^{\rm noise}_l)$, \citep{tegmark/deoliveira-costa:1998, refregier/spergel/herbig:2000},
where $b_l$ is the transfer function of the \map\ beam response \citep{jarosik/etal:prep},
$C^{\rm cmb}_l$ is the CMB angular power spectrum, and $C^{\rm noise}_l$ is the noise power.
The filtering suppresses CMB and Galactic foreground fluctuations relative to point
sources.  For peaks in the filtered maps that are $\gt 5\sigma$ in any band, the
unfiltered temperature maps are fit with the sum of a Gaussian profile and a planar baselevel.
The Gaussian amplitude is converted to a source flux density using the conversion
factors given in Table~2 of \citet{jarosik/etal:prep}, and flux density uncertainty is
calculated from the $1 \sigma$ uncertainty in the fit amplitude.  The source is entered
into the catalog if the fit source width is within a factor of two of the beam width.
Flux density values are entered for bands where they exceed $2\sigma$.  A point source
catalog mask is used to exclude sources in Galactic plane and Magellanic cloud regions.
This mask has changed from the five-year analysis.  A map pixel is outside of the new mask if
it is either outside of the diffuse component of the seven-year KQ85y7 temperature analysis mask
or outside of the five-year point source catalog mask.  This mask admits 82\% of the sky,
compared to 78\% for the five-year version.  We identify possible 5 GHz counterparts to the
\WMAP\ sources by cross-correlating with the GB6 \citep{gregory/etal:1996}, PMN 
\citep{griffith/etal:1994, griffith/etal:1995, wright.a/etal:1994, wright.a/etal:1996},
\citet{kuehr/etal:1981}, and \citet{healey/etal:2009} catalogs.  
A 5 GHz source is identified as a counterpart if it lies within $11\arcmin$ of
the \map\ source position (the mean \map\ source position uncertainty is $4\arcmin$, and can be twice as large for faint sources near the detection threshold).
When two or more 5 GHz sources are within $11\arcmin$, the brightest is assumed to be
the counterpart and a multiple identification flag is entered in the catalog.

The second method of point source identification is the CMB-free method originally applied
to one-year and three-year $V$- and $W$-band maps by \citet{chen/wright:2008} and to five-year $V$- and $W$-band maps by
\citet{wright/etal:2009}.  The method used here is that applied to five-year $Q$, $V$, and $W$ maps
by \citet{chen/wright:2009}.  The $V$- and $W$-band maps are smoothed to $Q$-band resolution.  A special
ILC map is then formed from the three maps using weights such
that CMB fluctuations are removed, flat-spectrum point sources are retained with fluxes
normalized to $Q$-band, and the variance of the ILC map is minimized.  The ILC map is filtered
to reduce the noise and suppress large angular scale structure.  Peaks in the filtered map
that are $\gt 5\sigma$ and outside of the seven-year point source catalog mask are identified as
point sources, and source positions are obtained by fitting the beam profile plus a baseline to
the filtered map for each source.  Source fluxes are estimated by integrating the $Q$, $V$, and $W$
temperature maps within $1\degrees.25$ of each source position, with a weighting
function to enhance the contrast of the point source relative to background fluctuations,
and applying a correction for Eddington bias due to noise.  Detected sources were identified
with sources in the five-year \map\ five-band catalog \citep{wright/etal:2009} and the five-year QVW
catalog \citet{chen/wright:2009} if the positions agreed within $15\arcmin$.  They were also
correlated against the 5GHz GB6, PMN, and \citet{kuehr/etal:1981} catalogs to identify 
possible 5 GHz counterparts within $15\arcmin$.  Optical identifications were made by
searching the NASA Extragalactic Database.

The seven-year five-band point source catalog is presented in Table~\ref{tab:fivebandptsrc} and the seven-year QVW point source
catalog is presented in Table~\ref{tab:qvwptsrc}.  The five-band catalog contains 471 sources, the QVW catalog
contains 417 sources, and the two catalogs have 346 sources in common.  For comparison, the
five-year five-band catalog contained 390 sources, the five-year QVW catalog contained 381 sources,
and they had 287 sources in common.  Differences in the source populations detected by the
two search methods do not appear to be mainly due to spectral index differences.  The distribution
of spectral index in the five \map\ bands for the sources that are only in the five-band catalog is
similar to that for the sources common to both catalogs.  The differences are thought to be largely
caused by Eddington bias in the five-band source detections due to CMB fluctuations and noise.  At low
flux levels, the five-band method tends to detect point sources located on positive CMB fluctuations
and to overestimate their fluxes, and it tends to miss sources located in negative CMB fluctuations.
This was shown by application of the method to simulated skymaps \citep{wright/etal:2009}, and its
effect is also seen in the comparison by \citet{chen/wright:2009} of five-year fluxes from the five-band
method with those from the CMB-free method in $Q$-, $V$-, and $W$-bands.  

%%%%%
% Conc
%%%%%
\section{Conclusions\label{sec:conclusions}}

Even with all the uncertainty regarding foregrounds in the Galactic plane, we find no evidence for foreground contamination outside our current KQ85y7 analysis mask.  Further, the cleaning process requires only three simple power-law foregrounds, and leaves no more than 15 $\mu\textrm{K}$ of residuals in the CMB temperature power spectrum.

We find no evidence of polarized foregrounds beyond those from soft (steep-spectrum) synchrotron and thermal dust emission.  In particular, we see no indication of an energetic population of synchrotron-emitting electrons near the Galactic center.

Additional years of data have allowed us to detect a combined 62 new point sources using two  techniques, a 12\% increase from the five-year data release.  A total of 346 point sources are in common between the two techniques.

More and more evidence is indicating that within a few degrees of the Galactic plane, the behavior of low-frequency foregrounds is complicated and has not been completely understood.  \WMAP\ data show a rapidly steepening spectrum from 20-40 GHz, which may be explained as emission from spinning dust grains.  The leading systematic, beam asymmetry, does not appear able to alter the spectrum enough to eliminate the need for spinning dust or a similar component.
ARCADE data appear consistent with the spinning dust explanation, although some discrepancies remain as to the relative strength of the emission.  More data at frequencies where  spinning dust emission is expected to be strongest (10-40 GHz) would be very helpful.

\acknowledgements
The \WMAP\ mission is made possible by the support of the Science Mission Directorate Office at NASA Headquarters.  This research was additionally supported by NASA grants NNG05GE76G, NNX07AL75G S01, LTSA03-000-0090, ATPNNG04GK55G, and ADP03-0000-092.  This research has made use of NASA's Astrophysics Data System Bibliographic Services.  We acknowledge use of the HEALPix, CAMB, and CMBFAST packages.

%%%% TABLES %%%%%%%%%%%%%%%%%%%%%%%%%%%%%%%%
\begin{deluxetable}{crrrrr}
      %WMAP% from /map/flight4/pass4/components/clean_bicg_maps/y7v1_plancut_new_final/ilc/
      \tablewidth{0pt}
      \tablecaption{ILC Coefficients Per Region\tablenotemark{a}\label{tab:ilc}}
      \startdata
      \hline\hline
      Region & 
       	\multicolumn{1}{c}{$K$-band} &
	\multicolumn{1}{c}{$Ka$-band}   &
	\multicolumn{1}{c}{$Q$-band}      &
	\multicolumn{1}{c}{$V$-band}  &
	\multicolumn{1}{c}{$W$-band}\\
      \hline
   0  &  0.1495  & -0.7184  & -0.3188  &  2.3071  & -0.4195 \\
   1  & -0.0035  & -0.2968  & -0.1963  &  2.0533  & -0.5567 \\
   2  &  0.0258  & -0.3368  & -0.3162  &  1.8368  & -0.2096 \\
   3  & -0.0945  &  0.1772  & -0.6087  &  1.5541  & -0.0281 \\
   4  & -0.0771  &  0.0881  & -0.4149  &  0.9559  &  0.4480 \\
   5  &  0.1928  & -0.7451  & -0.4538  &  2.4673  & -0.4612 \\
   6  & -0.0918  &  0.1946  & -0.5586  &  1.0227  &  0.4332 \\
   7  &  0.1533  & -0.7464  & -0.2033  &  2.2798  & -0.4834 \\
   8  &  0.2061  & -0.2979  & -1.5705  &  3.5678  & -0.9056 \\
   9  & -0.0889  & -0.1241  & -0.0816  &  1.2066  &  0.0880 \\
  10  &  0.1701  & -0.8610  & -0.1825  &  2.8264  & -0.9530 \\
  11  &  0.2358  & -0.8467  & -0.6020  &  2.8336  & -0.6206 \\
      \enddata
      \tablenotetext{a}{The ILC temperature (in thermodynamic units) at pixel $p$ of region $n$ is $T_n(p) = \sum_{i=1}^5 \zeta_{n,i} T^i(p)$, where $\zeta$ are the coefficients above and the sum is over \WMAP's frequency bands.}
\end{deluxetable}

\begin{deluxetable}{cccccc}
      \tablewidth{0pt}
      \tablecaption{Template Cleaning Temperature Coefficients \label{tab:template}}
      \startdata
      \hline\hline
      DA\tablenotemark{a} & $b_1$ & $b_2$ ($\mu$K R$^{-1}$) & $b_3$ & $\beta_s \tablenotemark{b} $ & $h_\mathrm{ff}$\tablenotemark{c}($\mu$K R$^{-1}$)\\
      \hline
      Q1 &  0.234  & 1.206 &  0.203 & -3.26  & 7.12\\
      Q2  & 0.232  & 1.240 &  0.201 & -3.30  & 7.13\\
      V1 &  0.048 &  0.791 &  0.466 & -3.63  & 7.20\\
   V2  & 0.045 &  0.772  & 0.483 & -3.64  & 7.21\\
   W1 &  0.000 &  0.436  & 1.277 & \nodata  & 7.24\\
   W2 &  0.000 &  0.430  & 1.291 & \nodata  & 7.24\\
   W3 &  0.000 &  0.438  & 1.257 & \nodata  & 7.24\\
   W4 &  0.000 &  0.432  & 1.285 & \nodata &  7.24\\
      \enddata
      \tablenotetext{a}{\WMAP\ has two differencing assemblies (DAs) for $Q$- and $V$-bands and four for $W$-band; the high signal-to-noise in total intensity allows each DA to be fitted independently.}
      \tablenotetext{b}{Power law slope relative to $K$-band, as derived from $b_1$; $W$-band values are less than -4.}
      \tablenotetext{c}{Free-free to H$\alpha$ ratio at $K$-band, as derived from $b_1$ and $b_2$.  The expected value for an electron temperature of 8000 K is 11.4 $\mu$K R$^{-1}$ \citep{bennett/etal:2003c}.}
\end{deluxetable}

\begin{deluxetable}{ccccc}
\tablewidth{0pt}
\tablecaption{Template Cleaning Polarization Coefficients\label{tab:poltemplate}}
\startdata
\hline\hline
Band & $a_1$\tablenotemark{a} & $\beta_s(\nu_{K},\nu)$\tablenotemark{b} &
$a_2$\tablenotemark{a} & $\beta_d(\nu,\nu_{W})\tablenotemark{b}$ \\
\hline
Ka &   0.3202  &   -3.13  &  0.0144 &    1.43 \\
Q  &  0.1683  &  -3.13 &  0.0177    &   1.54 \\
V  &  0.0613  & -2.93   &  0.0355  &   1.50 \\
W  &  0.0412   &  -2.41  &  0.0770 &   \nodata \\
\enddata
\tablenotetext{a}{The $a_i$ coefficients are dimensionless and produce model maps from templates.%, both in thermodynamic temperature units.
}
\tablenotetext{b}{The spectral indices refer to antenna temperature.}
\end{deluxetable}

\begin{deluxetable}{lccc}
\tablewidth{0pt}
\tablecaption{Foreground Power Spectrum Parameters\label{tab:fgpowerspec}}
\startdata
\hline\hline
Component & ${\cal B}_c$ $[\mu\textrm{K}^2]$\tablenotemark{a} & $m$ \tablenotemark{a}& $N$ $[\mu\textrm{K}]$\tablenotemark{a}\\
\hline
Synchrotron EE & $271\pm 31$ & $-0.73\pm 0.04$ & $0.109\pm 0.001$ \\
Synchrotron BB & $130\pm 8.6$ & $-0.61\pm 0.02$ & $0.107\pm 0.001$ \\
Dust EE  & $17.7\pm 2.5$ & $-1.13\pm 0.06$ & $0.065 \pm 0.001$ \\
Dust BB & $6.41 \pm 1.1$ & $-0.65\pm 0.06$ & $0.066 \pm 0.001$ \\
\enddata
\tablenotetext{a}{Quoted errors are only statistical uncertainty from the fitting process.}
\end{deluxetable}

%%%%%%%%%%%%%%%%%%%%%%%%%%%%%%%%%%%%%%%%%
%%%% FIGURES %%%%%%%%%%%%%%%%%%%%%%%%%%%%%%%

%%%%
%%%% point source tables %%%%%%%%%%%%%%%%%%%%%%%%%
%%%%

\begin{deluxetable}{ccccccccccc}
\rotate
\tablecaption{\label{tab:fivebandptsrc} \WMAP\ Five-Band Point Source Catalog}
\tabletypesize{\footnotesize}
\tablewidth{0pt}
\tablehead{
 \colhead{RA [hms]}
&\colhead{Dec [dm]}
&\colhead{ID}
&\colhead{$K$ [Jy]}
&\colhead{$Ka$ [Jy]}
&\colhead{$Q$ [Jy]}
&\colhead{$V$ [Jy]}
&\colhead{$W$ [Jy]}
&\colhead{$\alpha$}
&\colhead{5 GHz ID}
&
}
\startdata
 00 03 30
&$-$47 50
&   
&$0.7\pm0.06$
&$0.7\pm0.05$
&$0.7\pm0.07$
&\nodata
&\nodata
&$0.1\pm0.5$
&\nodata
&
\\
 00 06 07
&$-$06 23
&060
&$2.3\pm0.06$
&$2.1\pm0.1$
&$2.1\pm0.1$
&$1.9\pm0.2$
&\nodata
&$-0.2\pm0.2$
&PMN J0006-0623
&
\\
 00 10 37
&\phs11 01
&   
&$0.9\pm0.07$
&$1.0\pm0.09$
&$1.0\pm0.1$
&$1.2\pm0.2$
&$1.2\pm0.3$
&$0.2\pm0.3$
&GB6 J0010+1058
&
\\
 00 12 47
&$-$39 52
&202
&$1.3\pm0.04$
&$1.2\pm0.07$
&$0.9\pm0.07$
&$0.8\pm0.1$
&$0.7\pm0.2$
&$-0.4\pm0.2$
&PMN J0013-3954
&
\\
 00 19 12
&\phs26 01
&   
&$0.7\pm0.06$
&$0.5\pm0.09$
&$0.7\pm0.1$
&$0.5\pm0.1$
&\nodata
&$-0.3\pm0.5$
&GB6 J0019+2602
&
\\
 00 19 40
&\phs20 20
&   
&$1.0\pm0.05$
&$1.1\pm0.07$
&$0.9\pm0.08$
&$1.1\pm0.2$
&\nodata
&$0.0\pm0.3$
&GB6 J0019+2021
&
\\
 00 25 22
&$-$26 03
&   
&$0.9\pm0.05$
&$0.7\pm0.07$
&$0.5\pm0.08$
&\nodata
&\nodata
&$-0.8\pm0.5$
&PMN J0025-2602
&\tablenotemark{a}
\\
 00 26 06
&$-$35 10
&   
&$1.1\pm0.06$
&$1.2\pm0.08$
&$1.4\pm0.1$
&$1.1\pm0.1$
&$0.5\pm0.2$
&$0.0\pm0.3$
&PMN J0026-3512
&
\\
\enddata
\tablenotetext{a}{Indicates the source has multiple possible identifications.}
\tablenotetext{b}{Source J0322-3711 (Fornax A) is extended, and the fluxes listed were obtained by aperture photometry.}
\tablenotetext{c}{Source J0519-0540 is a blend of the Lynds Bright Nebulae LBN 207.65-23.11 and LBN 207.29-22.66.}
\tablenotetext{d}{Source J1356+7645 is outside of the declination range of the GB6 and PMN catalogs.  
Identified as QSO NVSSJ135755+764320 by Trushkin (2006, private communication).}
\tablenotetext{e}{Source J1633+8227 is outside of the declination range of the
GB6 and PMN catalogs. It was identified as NGC 6251 by \citet{trushkin:2003}.}
\end{deluxetable}

\begin{deluxetable}{lccccccccc}
%need rotate even if scriptsize
\rotate
\tablecolumns{10}
%\tabletypesize{\scriptsize}
\tabletypesize{\footnotesize}
\tablewidth{0pt}
\tablecaption{\label{tab:qvwptsrc}WMAP CMB-Free QVW Point Source Catalog} 
\tablehead{
\colhead{RA} & \colhead{DEC}  &
\colhead{WMAP/QVW5} & \colhead{Type} & \colhead{Q} & \colhead{V} & \colhead{W}& \colhead{5 GHz ID} & \colhead{Dist.}  & \colhead{Note } \\
\colhead{dms} & \colhead{dms} & \colhead{ID} &
\colhead{}  & \colhead{[Jy]} & \colhead{[Jy]}  & \colhead{[Jy]} &  \colhead{} & \colhead{[arcmin]} & \colhead{} 
}
\startdata 
\hline
$ $00 06 11.5 & $-$06 24 57.6 & J0006-0623     & G       & $   2.2\pm   0.2$ & $   1.9\pm   0.2$ & $   1.2\pm   0.4$ & PMN J0006-0623      &    1.4 &   \\
$ $00 10 30.2 & $ $10 58 22.8 & J0010+1101     & G       & $   0.9\pm   0.2$ & $   1.1\pm   0.2$ & $   1.1\pm   0.4$ & GB6 J0010+1058      &    0.4 &   \\
$ $00 12 57.6 & $-$39 53 49.2 & J0012-3952     & RadioS  & $   0.8\pm   0.2$ & $   0.8\pm   0.2$ & $   1.0\pm   0.3$ & PMN J0013-3954      &    0.8 &   \\
$ $00 13 44.2 & $ $40 54 21.6 & QVW J0013+4051 & G       & $   0.6\pm   0.2$ & $   0.6\pm   0.2$ & $   0.5\pm   0.4$ & GB6 J0013+4051      &    3.8 &   \\
$ $00 19 40.1 & $ $25 59 31.2 & J0019+2603     & QSO     & $   0.9\pm   0.2$ & $   0.3\pm   0.2$ & $   0.5\pm   0.4$ & GB6 J0019+2602      &    3.3 &   \\
$ $00 26 04.6 & $-$35 12 39.6 & J0026-3510     & RadioS  & $   1.4\pm   0.2$ & $   1.0\pm   0.2$ & $   0.6\pm   0.4$ & PMN J0026-3512      &    1.6 &   \\
$ $00 29 49.7 & $ $05 53 38.4 & J0029+0554     & QSO     & $   0.9\pm   0.2$ & $   0.6\pm   0.2$ & $   0.8\pm   0.4$ & GB6 J0029+0554B     &    1.4 & a \\
$ $00 38 10.6 & $-$02 04 01.2 & QVW J0037-0207 & G       & $   0.7\pm   0.2$ & $   0.5\pm   0.2$ & $   0.7\pm   0.4$ & PMN J0038-0207      &    4.3 &   \\
\enddata

\tablenotetext{a}{Indicates this source has multiple identifications. The source listed here is the brighter one or the one with small offset when fluxes are comparable.}
\tablenotetext{b}{The source is associated with a different 5 GHz source than noted in the WMAP five-year point source catalog.}
\tablenotetext{c}{Indicates the source is in the three-year WMAP point source catalog, but not in the five-year catalog.}
\tablenotetext{d}{The source position is fitted to the local maximum of a 5 by 5 pixel neighborhood as there are two bright pixels within one $Q$-band beam, both of which are the brightest in their 3 by 3 pixel neighborhood. }
\tablenotetext{e}{Indicates no 5 GHz counterpart could be identified for the source.}

\end{deluxetable}

\clearpage

%%% appendices %%%%%%%%%%%%%%%%%%%%%%%%%%%%%%%%%%%%%

%The Reference List:
\bibliographystyle{apj}
\bibliography{apj-jour,temporary,wmap}

\begin{thebibliography}{51}
\expandafter\ifx\csname natexlab\endcsname\relax\def\natexlab#1{#1}\fi

\bibitem[{{Bennett} {et~al.}(2003){Bennett}, {Hill}, {Hinshaw}, {Nolta},
  {Odegard}, {Page}, {Spergel}, {Weiland}, {Wright}, {Halpern}, {Jarosik},
  {Kogut}, {Limon}, {Meyer}, {Tucker}, \& {Wollack}}]{bennett/etal:2003c}
{Bennett}, C.~L., {et~al.} 2003, \apjs, 148, 97

\bibitem[{{Bennett} {et~al.}(2010)}]{bennett/etal:prep}
---. 2010, in preparation

\bibitem[{{Chen} \& {Wright}(2008)}]{chen/wright:2008}
{Chen}, X., \& {Wright}, E.~L. 2008, \apj, 681, 747

\bibitem[{{Chen} \& {Wright}(2009)}]{chen/wright:2009}
---. 2009, \apj, 694, 222

\bibitem[{{Culverhouse} {et~al.}(2010){Culverhouse}, {Ade}, {Bock}, {Bowden},
  {Brown}, {Cahill}, {Castro}, {Church}, {Friedman}, {Ganga}, {Gear}, {Gupta},
  {Hinderks}, {Kovac}, {Lange}, {Leitch}, {Melhuish}, {Memari}, {Murphy},
  {Orlando}, {Schwarz}, {O' Sullivan}, {Piccirillo}, {Pryke}, {Rajguru},
  {Rusholme}, {Taylor}, {Thompson}, {Turner}, {Wu}, \&
  {Zemcov}}]{culverhouse/etal:2010}
{Culverhouse}, T., {et~al.} 2010, Astrophys. J., 722, 1057

\bibitem[{{Dobler} \&
  {Finkbeiner}(2008{\natexlab{a}})}]{dobler/finkbeiner:2008a}
{Dobler}, G., \& {Finkbeiner}, D.~P. 2008{\natexlab{a}}, \apj, 680, 1222

\bibitem[{{Dobler} \&
  {Finkbeiner}(2008{\natexlab{b}})}]{dobler/finkbeiner:2008b}
---. 2008{\natexlab{b}}, \apj, 680, 1235

\bibitem[{{Dobler} {et~al.}(2010){Dobler}, {Finkbeiner}, {Cholis}, {Slatyer},
  \& {Weiner}}]{dobler/etal:2009}
{Dobler}, G., {Finkbeiner}, D.~P., {Cholis}, I., {Slatyer}, T.~R., \& {Weiner},
  N. 2010, Astrophys. J., 717, 825

\bibitem[{{Draine} \& {Lazarian}(1998{\natexlab{a}})}]{draine/lazarian:1998a}
{Draine}, B.~T., \& {Lazarian}, A. 1998{\natexlab{a}}, \apjl, 494, L19

\bibitem[{{Draine} \& {Lazarian}(1998{\natexlab{b}})}]{draine/lazarian:1998b}
---. 1998{\natexlab{b}}, \apj, 508, 157

\bibitem[{{Draine} \& {Lazarian}(1999)}]{draine/lazarian:1999}
---. 1999, \apj, 512, 740

\bibitem[{{Dunkley} {et~al.}(2009{\natexlab{a}}){Dunkley}, {Komatsu}, {Nolta},
  {Spergel}, {Larson}, {Hinshaw}, {Page}, {Bennett}, {Gold}, {Jarosik},
  {Weiland}, {Halpern}, {Hill}, {Kogut}, {Limon}, {Meyer}, {Tucker}, {Wollack},
  \& {Wright}}]{dunkley/etal:2009}
{Dunkley}, J., {et~al.} 2009{\natexlab{a}}, \apjs, 180, 306

\bibitem[{{Dunkley} {et~al.}(2009{\natexlab{b}}){Dunkley}, {Amblard},
  {Baccigalupi}, {Betoule}, {Chuss}, {Cooray}, {Delabrouille}, {Dickinson},
  {Dobler}, {Dotson}, {Eriksen}, {Finkbeiner}, {Fixsen}, {Fosalba}, {Fraisse},
  {Hirata}, {Kogut}, {Kristiansen}, {Lawrence}, {Magalha\~{}Es},
  {Miville-Deschenes}, {Meyer}, {Miller}, {Naess}, {Page}, {Peiris},
  {Phillips}, {Pierpaoli}, {Rocha}, {Vaillancourt}, \&
  {Verde}}]{dunkley/etal:2009c}
{Dunkley}, J., {et~al.} 2009{\natexlab{b}}, in American Institute of Physics
  Conference Series, Vol. 1141, American Institute of Physics Conference
  Series, ed. {S.~Dodelson, D.~Baumann, A.~Cooray, J.~Dunkley, A.~Fraisse,
  M.~G.~Jackson, A.~Kogut, L.~Krauss, M.~Zaldarriaga, \& K.~Smith }, 222--264

\bibitem[{{Eriksen} {et~al.}(2007){Eriksen}, {Huey}, {Saha}, {Hansen}, {Dick},
  {Banday}, {G{\'o}rski}, {Jain}, {Jewell}, {Knox}, {Larson}, {O'Dwyer},
  {Souradeep}, \& {Wandelt}}]{eriksen/etal:2007}
{Eriksen}, H.~K., {et~al.} 2007, \apj, 656, 641

\bibitem[{{Finkbeiner}(2003)}]{finkbeiner:2003}
{Finkbeiner}, D.~P. 2003, \apjs, 146, 407, accepted (astro-ph/0301558)

\bibitem[{{Finkbeiner}(2004)}]{finkbeiner:2004}
---. 2004, \apj, 614, 186

\bibitem[{Finkbeiner {et~al.}(1999)Finkbeiner, Davis, \&
  Schlegel}]{finkbeiner/davis/schlegel:1999}
Finkbeiner, D.~P., Davis, M., \& Schlegel, D.~J. 1999, \apj, 524, 867

\bibitem[{{Fixsen} {et~al.}(2009){Fixsen}, {Kogut}, {Levin}, {Limon}, {Lubin},
  {Mirel}, {Seiffert}, {Singal}, {Wollack}, {Villela}, \&
  {Wuensche}}]{fixsen/etal:2010}
{Fixsen}, D.~J., {et~al.} 2009, \apj, submitted (arXiv:0901.0555)

\bibitem[{{Gold} {et~al.}(2009){Gold}, {Bennett}, {Hill}, {Hinshaw}, {Odegard},
  {Spergel}, {Weiland}, {Dunkley}, {Halpern}, {Jarosik}, {Kogut}, {Komatsu},
  {Larson}, {Meyer}, {Nolta}, {Wollack}, \& {Wright}}]{gold/etal:2009}
{Gold}, B., {et~al.} 2009, \apjs, 180, 265

\bibitem[{Gorski {et~al.}(2005)Gorski, Hivon, Banday, Wandelt, Hansen,
  Reinecke, \& Bartlemann}]{gorski/etal:2005}
Gorski, K.~M., Hivon, E., Banday, A.~J., Wandelt, B.~D., Hansen, F.~K.,
  Reinecke, M., \& Bartlemann, M. 2005, \apj, 622, 759

\bibitem[{{Gregory} {et~al.}(1996){Gregory}, {Scott}, {Douglas}, \&
  {Condon}}]{gregory/etal:1996}
{Gregory}, P.~C., {Scott}, W.~K., {Douglas}, K., \& {Condon}, J.~J. 1996,
  \apjs, 103, 427

\bibitem[{{Griffith} {et~al.}(1994){Griffith}, {Wright}, {Burke}, \&
  {Ekers}}]{griffith/etal:1994}
{Griffith}, M.~R., {Wright}, A.~E., {Burke}, B.~F., \& {Ekers}, R.~D. 1994,
  \apjs, 90, 179

\bibitem[{{Griffith} {et~al.}(1995){Griffith}, {Wright}, {Burke}, \&
  {Ekers}}]{griffith/etal:1995}
---. 1995, \apjs, 97, 347

\bibitem[{{Haslam} {et~al.}(1981){Haslam}, {Klein}, {Salter}, {Stoffel},
  {Wilson}, {Cleary}, {Cooke}, \& {Thomasson}}]{haslam/etal:1981}
{Haslam}, C.~G.~T., {Klein}, U., {Salter}, C.~J., {Stoffel}, H., {Wilson},
  W.~E., {Cleary}, M.~N., {Cooke}, D.~J., \& {Thomasson}, P. 1981, \aap, 100,
  209

\bibitem[{{Healey} {et~al.}(2009){Healey}, {Fuhrmann}, {Taylor}, {Romani}, \&
  {Readhead}}]{healey/etal:2009}
{Healey}, S.~E., {Fuhrmann}, L., {Taylor}, G.~B., {Romani}, R.~W., \&
  {Readhead}, A. C.~S. 2009, \aj, 138, 1032

\bibitem[{{Hinshaw} {et~al.}(2007){Hinshaw}, {Nolta}, {Bennett}, {Bean},
  {Dor{\'e}}, {Greason}, {Halpern}, {Hill}, {Jarosik}, {Kogut}, {Komatsu},
  {Limon}, {Odegard}, {Meyer}, {Page}, {Peiris}, {Spergel}, {Tucker}, {Verde},
  {Weiland}, {Wollack}, \& {Wright}}]{hinshaw/etal:2007}
{Hinshaw}, G., {et~al.} 2007, \apjs, 170, 288

\bibitem[{{Hooper} {et~al.}(2007){Hooper}, {Finkbeiner}, \&
  {Dobler}}]{hooper/finkbeiner/dobler:2007}
{Hooper}, D., {Finkbeiner}, D.~P., \& {Dobler}, G. 2007, \prd, 76, 083012

\bibitem[{{Jarosik} {et~al.}(2010)}]{jarosik/etal:prep}
{Jarosik}, N., {et~al.} 2010, in preparation

\bibitem[{{Kim} {et~al.}(2008){Kim}, {Naselsky}, \&
  {Christensen}}]{kim/naselsky/christensen:2008}
{Kim}, J., {Naselsky}, P., \& {Christensen}, P.~R. 2008, \prd, 77, 103002

\bibitem[{{Kogut} {et~al.}(2009){Kogut}, {Fixsen}, {Levin}, {Limon}, {Lubin},
  {Mirel}, {Seiffert}, {Singal}, {Villela}, {Wollack}, \&
  {Wuensche}}]{kogut/etal:2010}
{Kogut}, A., {et~al.} 2009, \apj, submitted (arXiv:0901.0562)

\bibitem[{{Komatsu} {et~al.}(2010)}]{komatsu/etal:prep}
{Komatsu}, E., {et~al.} 2010, in preparation

\bibitem[{{K\"uhr} {et~al.}(1981){K\"uhr}, {Witzel}, {Pauliny-Toth}, \&
  {Nauber}}]{kuehr/etal:1981}
{K\"uhr}, H., {Witzel}, A., {Pauliny-Toth}, I.~I.~K., \& {Nauber}, U. 1981,
  \aaps, 45, 367

\bibitem[{{Kunz} {et~al.}(2006){Kunz}, {Trotta}, \&
  {Parkinson}}]{kunz/trotta/parkinson:2006}
{Kunz}, M., {Trotta}, R., \& {Parkinson}, D.~R. 2006, \prd, 74, 023503

\bibitem[{{Larson} {et~al.}(2010)}]{larson/etal:prep}
{Larson}, D., {et~al.} 2010, in preparation

\bibitem[{{Lawson} {et~al.}(1987){Lawson}, {Mayer}, {Osborne}, \&
  {Parkinson}}]{lawson/etal:1987}
{Lawson}, K.~D., {Mayer}, C.~J., {Osborne}, J.~L., \& {Parkinson}, M.~L. 1987,
  \mnras, 225, 307

\bibitem[{{Lazarian} \& {Draine}(2000)}]{lazarian/draine:2000}
{Lazarian}, A., \& {Draine}, B.~T. 2000, \apjl, 536, L15

\bibitem[{{Lewis} {et~al.}(2000){Lewis}, {Challinor}, \&
  {Lasenby}}]{lewis/challinor/lasenby:2000}
{Lewis}, A., {Challinor}, A., \& {Lasenby}, A. 2000, \apj, 538, 473

\bibitem[{{Page} {et~al.}(2007){Page}, {Hinshaw}, {Komatsu}, {Nolta},
  {Spergel}, {Bennett}, {Barnes}, {Bean}, {Dor{\'e}}, {Dunkley}, {Halpern},
  {Hill}, {Jarosik}, {Kogut}, {Limon}, {Meyer}, {Odegard}, {Peiris}, {Tucker},
  {Verde}, {Weiland}, {Wollack}, \& {Wright}}]{page/etal:2007}
{Page}, L., {et~al.} 2007, \apjs, 170, 335

\bibitem[{{Refregier} {et~al.}(2000){Refregier}, {Spergel}, \&
  {Herbig}}]{refregier/spergel/herbig:2000}
{Refregier}, A., {Spergel}, D.~N., \& {Herbig}, T. 2000, \apj, 531, 31

\bibitem[{{Scaife} {et~al.}(2009){Scaife}, {Hurley-Walker}, {Green}, {Davies},
  {Franzen}, {Grainge}, {Hobson}, {Lasenby}, {Pooley}, {Rodriguez-Gonzalvez},
  {Saunders}, {Scott}, {Shimwell}, {Titterington}, {Waldram}, \&
  {Zwart}}]{scaife/etal:2010}
{Scaife}, T.~A.~C.~A.~M.~M., {et~al.} 2009, Mon. Not. Roy. Astron. Soc., 400,
  1394

\bibitem[{{Schlegel} {et~al.}(1998){Schlegel}, {Finkbeiner}, \&
  {Davis}}]{schlegel/finkbeiner/davis:1998}
{Schlegel}, D.~J., {Finkbeiner}, D.~P., \& {Davis}, M. 1998, \apj, 500, 525

\bibitem[{{Singal} {et~al.}(2009){Singal}, {Fixsen}, {Kogut}, {Levin}, {Limon},
  {Lubin}, {Mirel}, {Seiffert}, {Villela}, {Wollack}, \&
  {Wuensche}}]{singal/etal:2010}
{Singal}, J., {et~al.} 2009, \apj, submitted (arXiv:0901.0546)

\bibitem[{Tegmark \& de~Oliveira-Costa(1998)}]{tegmark/deoliveira-costa:1998}
Tegmark, M., \& de~Oliveira-Costa, A. 1998, \apjl, 500, L83

\bibitem[{{Tibbs} {et~al.}(2010){Tibbs}, {Watson}, {Dickinson}, {Davies},
  {Davis}, {del Burgo}, {Franzen}, {G{\'e}nova-Santos}, {Grainge}, {Hobson},
  {Padilla-Torres}, {Rebolo}, {Rubi{\~n}o-Mart{\'{\i}}n}, {Saunders}, {Scaife},
  \& {Scott}}]{tibbs/etal:2010}
{Tibbs}, C.~T., {et~al.} 2010, Mon. Not. Roy. Astron. Soc., 402, 1969

\bibitem[{Trushkin(2003)}]{trushkin:2003}
Trushkin, S.~A. 2003, Bull. Spec. Astrophys. Obs. N. Caucasus, 55, 90

\bibitem[{{Veneziani} {et~al.}(2010){Veneziani}, {Ade}, {Bock}, {Boscaleri},
  {Crill}, {de Bernardis}, {De Gasperis}, {de Oliveira-Costa}, {De Troia}, {Di
  Stefano}, {Ganga}, {Jones}, {Kisner}, {Lange}, {MacTavish}, {Masi},
  {Mauskopf}, {Montroy}, {Natoli}, {Netterfield}, {Pascale}, {Piacentini},
  {Pietrobon}, {Polenta}, {Ricciardi}, {Romeo}, \&
  {Ruhl}}]{veneziani/etal:2009}
{Veneziani}, M., {et~al.} 2010, Astrophys. J., 713, 959

\bibitem[{{Vio} \& {Andreani}(2009)}]{vio/andreani:2009}
{Vio}, R., \& {Andreani}, P. 2009, arXiv:0910.4294

\bibitem[{{Weiland} {et~al.}(2010)}]{weiland/etal:prep}
{Weiland}, J.~L., {et~al.} 2010, in preparation

\bibitem[{{Wright} {et~al.}(1994){Wright}, {Griffith}, {Burke}, \&
  {Ekers}}]{wright.a/etal:1994}
{Wright}, A.~E., {Griffith}, M.~R., {Burke}, B.~F., \& {Ekers}, R.~D. 1994,
  \apjs, 91, 111

\bibitem[{{Wright} {et~al.}(1996){Wright}, {Griffith}, {Hunt}, {Troup},
  {Burke}, \& {Ekers}}]{wright.a/etal:1996}
{Wright}, A.~E., {Griffith}, M.~R., {Hunt}, A.~J., {Troup}, E., {Burke}, B.~F.,
  \& {Ekers}, R.~D. 1996, \apjs, 103, 145

\bibitem[{{Wright} {et~al.}(2009){Wright}, {Chen}, {Odegard}, {Bennett},
  {Hill}, {Hinshaw}, {Jarosik}, {Komatsu}, {Nolta}, {Page}, {Spergel},
  {Weiland}, {Wollack}, {Dunkley}, {Gold}, {Halpern}, {Kogut}, {Larson},
  {Limon}, {Meyer}, \& {Tucker}}]{wright/etal:2009}
{Wright}, E.~L., {et~al.} 2009, \apjs, 180, 283

\end{thebibliography}

%\begin{thebibliography}{50}
%%\expandafter\ifx\csname natexlab\endcsname\relax\def\natexlab#1{#1}\fi
%\end{thebibliography}

\end{document}